\documentclass[12pt,preprint]{aastex}

\newcommand{\JHKs} {JHK_S}
\newcommand{\msun} {$M_{\sun}$}

\newcommand{\Te} {T_{\rm eff}}
\newcommand{\logg} {\log g} 
\begin{document}

\title{Infrared Photometric Analysis of White Dwarfs from The Two 
Micron All Sky Survey and the Spitzer Space Telescope}

\author{P.-E. Tremblay and P. Bergeron}
\affil{D\'epartement de Physique, Universit\'e de Montr\'eal, C.P.~6128, 
Succ.~Centre-Ville, Montr\'eal, Qu\'ebec, Canada, H3C 3J7.}
\email{tremblay@astro.umontreal.ca, bergeron@astro.umontreal.ca}

\begin{abstract}

We review the available near- and mid-infrared photometry for white
dwarfs obtained from the Two Micron All-Sky Survey (2MASS) and by the
{\it Spitzer Space Telescope}.  Both data sets have recently been used
to seek white dwarfs with infrared excesses due to the presence of
unresolved companions or circumstellar disks, and also to derive the
atmospheric parameters of cool white dwarfs.  We first attempt to
evaluate the reliability of the 2MASS photometry by comparing it with
an independent set of published $JHK$ CIT magnitudes for 160 cool
white dwarf stars, and also by comparing the data with the predictions
of detailed model atmosphere calculations.  The possibility of using
2MASS to identify unresolved M dwarf companions or circumstellar disks
is then discussed.  We also revisit the analysis of 46 binary
candidates from Wachter et al.~using the synthetic flux method and
confirm the large near-infrared excesses in most objects. We perform a
similar analysis by fitting {\it Spitzer} 4.5 and 8 $\mu$m photometric
observations of white dwarfs with our grid of model atmospheres, and
demonstrate the reliability of both the {\it Spitzer} data and the
theoretical calculations up to 8 $\mu$m. Finally, we search for
massive disks resulting from the merger of two white dwarfs in a 2MASS
sample composed of 57 massive degenerates, and show that massive disks
are uncommon in such stars.

\end{abstract}


\section{INTRODUCTION}

With the recent All-Sky Data Release of the Two Micron All-Sky
Survey\footnote{See
http://www.ipac.caltech.edu/2mass/releases/allsky/} (2MASS), we are
now able to retrieve near-infrared (NIR) $J$, $H$, and $K_S$
magnitudes for more than a thousand white dwarfs that fall within the
2MASS detection limit. This database was used in several studies aimed
at identifying new cool white dwarfs \citep[e.g.,][]{marcos05} or
circumstellar disks \citep{kilic06a} and seeking binary candidates
\citep{wachter03,holberg05,debes05}. In the latter case, one of the
main interests are the binary systems containing a main sequence star
and a white dwarf. These systems might reveal important details about
stellar populations and evolution. Different techniques have been used
to seek these binary candidates. Until recently, most systematic
searches were based on surveys of resolved common proper-motion
binaries \citep{silvestri02}, but new interest has emerged for
identifying unresolved binaries. One of the reasons is that accretion
from a previously unknown close companion could account for the high
metal abundances observed in some white dwarfs. The preferred method
for seeking unresolved binary candidates is to perform a photometric
analysis. In the case where the companion is an M dwarf, the white
dwarf star usually dominates the observed flux in the optical
regions. Therefore, it is natural to look for an excess in the NIR,
either photometrically or spectroscopically, where the contribution
from the M dwarf becomes dominant \citep[see][for a review]{dobbie05}.

Exploiting the 2MASS photometric data, different methods of analysis
were used to identify NIR excesses. \citet{wachter03} used the
second incremental 2MASS data release, which covers about 50\% of the
sky. The authors took the approach of a $(J-H,H-K_S)$ two-color
diagram for 795 white dwarfs recovered from the 2MASS survey. They
identified 95 binary candidates, including 47 objects with prior
evidence of binarity. They also suggested 15 additional tentative
binary candidates. \citet{wellhouse05} used a similar two-color
diagram approach with a sample of 51 magnetic white dwarfs as
candidates for potential pre-cataclysmic variables. While they did not
find any binary candidates, they identified 10 objects with peculiar
colors associated with very low mass companions or
debris. \citet{holberg05} used the final 2MASS All-Sky Data Release to
study the 347 DA stars from the Palomar-Green Survey
\citep{liebert05}. Their technique relies on the spectroscopic
determinations of effective temperature and surface gravity, which
combined with the observed $V$ magnitude, can be used to compare
magnitudes predicted at $J$, $H$, and $K_S$ with those available in
the 2MASS Point-Source Catalog (PSC). The same technique had been used
before by
\citet{zuckerman92} and \citet{green00} but with independent NIR
photometric data sets. The disadvantage of this technique is that
reliable atmospheric parameters and $V$ magnitudes must
be available for each star.

As the low-mass main-sequence companion gets cooler --- typical of
late-type M or L dwarfs --- only a mild NIR excess is observed. The
NIR excesses expected from circumstellar dust disks and planets around
white dwarfs could be even less significant. \citet{zuckerman87} were
the first to identify such a system for the 0.7 \msun\ DAZ star G29-38
(2326$+$049), also a ZZ Ceti pulsator. More recently, \citet{kilic05}
and \citet{becklin05} went through a detailed analysis of GD~362
(1729$+$371), a massive DAZ star with unusually high metal abundances,
some nearly solar
\citep{gianninas04}. For both objects, there was a small but
significant excess in the NIR that could be detected in the $K$
band. However, it is from the large mid-infrared (MIR) excess
\citep{reach05,becklin05} that the disks could be
confirmed. NIR spectroscopic observations and 2MASS data have also been used by
\citet{kilic06a} to identify a third DAZ white dwarf, GD~56, that could harbor a
circumstellar disk, although this object has yet to be observed in the
MIR. \citet{chary99} and \citet{kilic05,kilic06a} analyzed a dozen
other DA and DAZ stars and found no evidence for similar circumstellar
disks. \citet{jura03} discussed possible scenarios and concluded that
not all white dwarfs with heavy elements in their atmospheres possess
a dust disk similar to that of G29-38. The current picture is that as
much as 14\% of the DAZ stars host a circumstellar disk
\citep{kilic06a}.

According to \citet{livio05}, disks and planets could also result from
the merger of two white dwarfs. Hence, the high-mass tail of the white
dwarf mass distribution \citep[see, e.g.,][]{liebert05} would
represent the most promising candidates to search for such disks or
planets. Livio et al.~suggest that a typical dust disk would have a
mass and radius of $M_d\sim0.007$
\msun\ and $R_d\sim 1$ AU, respectively. This is much larger and
massive than the disk proposed for G29-38
\citep{jura03}. Therefore, the predicted flux excess should be easily
detected in the NIR (assuming a standard composition and geometry)
and the 2MASS survey should provide a useful tool to further constrain
the proposed model.

In addition to the 2MASS NIR photometry, there is a developing
interest to observe white dwarfs at longer wavelengths in the
MIR. The {\it Spitzer Space Telescope} IRAC\footnote{See
http://ssc.spitzer.caltech.edu/irac/} photometry and IRS infrared
spectroscopy have been used in recent surveys of relatively bright,
nearby white dwarfs to better constrain the atmospheric parameters of
cool white dwarfs \citep{kilic06b} and to seek MIR excesses from disks
\citep{reach05,hansen06}. Since the contribution of a cold disk
becomes dominant only in the MIR, the {\it Spitzer} data set is more
sensitive to search for disks than the NIR 2MASS data set.

Before undertaking a more systematic search of white dwarf stars in
binaries or of circumstellar disk systems using 2MASS or {\it Spitzer}
data, it seems appropriate as a first step to evaluate properly the
reliability of the infrared photometric data sets and the ability of
current model atmospheres to reproduce the observations. We thus
present in \S~2 a comparison of 2MASS photometry with published $JHK$
magnitudes on the CIT photometric system for 160 cool white dwarfs,
and assess the limitations of the 2MASS survey. We then evaluate in
\S~3 the usefulness of the 2MASS photometric data for identifying
binary candidates using various techniques, and discuss the
implications of our results on several studies published in the
literature.  In \S~4, we perform a similar analysis but using the {\it
Spitzer} IRAC 4.5 and $8~\mu$m photometry presented in
\citet{kilic06b}.  Finally in \S~5, we analyze a sample of 57 white
dwarfs with spectroscopic masses above 0.8 \msun\ together with 2MASS
photometry to search for disks around massive white dwarfs, such as
those predicted by \citet{livio05}. Our conclusions follow in \S~6.

\section{COMPARISON OF CIT AND 2MASS PHOTOMETRY}

Our photometric sample used to compare against the 2MASS data is drawn
from the detailed photometric and spectroscopic analyses of
\citet[][hereafter BRL97]{bergeron97}, \citet[][]{leggett98}, 
and \citet[][hereafter BLR01]{bergeron01} who obtained improved
atmospheric parameters of cool white dwarfs from a comparison of
optical $BVRI$ and infrared $JHK$ photometry with the predictions of
model atmospheres appropriate for these stars. We selected from these
studies 183 cool white dwarfs with infrared $JHK$ magnitudes measured
on the CIT photometric system (with the exception of 0704$-$508
that has no $K$ measurement). This sample covers a range of effective
temperatures between $\Te\sim 4000$~K and 13,000~K, and all objects
have been successfully fitted by BRL97 and BLR01 under the assumption of
single stars (or double degenerates) with no evidence for any infrared
excess that could be due to the presence of an unresolved low-mass
main sequence star.

We searched the 2MASS PSC for all white dwarfs in our sample using the
GATOR batch file tool and a $20''$ search window centered on a set of
improved coordinates measured by J.~B.~Holberg (2005, private
communication). In most instances, multiple sources were found within
the search window and we unambiguously identified each object by
comparing the 2MASS atlas with the finding charts available from the
online version of the Villanova White Dwarf
Catalog\footnote{http://www.astronomy.villanova.edu/WDCatalog/index.html}.
We recovered the 2MASS $J$, $H$, and $K_S$ magnitudes for 160 stars
from our initial CIT photometric sample of 183 objects. The remaining
23 objects were dropped from our analysis for the following reasons: 9
were too faint for the 2MASS survey, 11 were not properly resolved due
to the presence of a nearby star, and 3 could not be unambiguously
identified from the comparison of the 2MASS atlas and the published
finding charts.  Our final sample of 160 cool white dwarfs is
presented in Table 1 where we provide the CIT and 2MASS magnitudes
for each object.  The uncertainties of the CIT magnitudes are 5\%
except where noted in Table 1, and the 2MASS photometric uncertainties
are given in parentheses (magnitudes with null uncertainties represent
lower limits).

Since the two data sets rely on completely different photometric
systems, we must keep in mind that there could be a possible offset
between both systems. For instance, \citet{carpenter01} have obtained
an empirical color transformation (see their eqs.~12 to 15) based on a
comparison of CIT and 2MASS photometry for 41 stars.
However, since this transformation has been obtained in a broad
general context and not specifically for cool white dwarfs, we 
first compare directly both photometric data sets without any
transformation, and discuss the possible offsets in the present context.

Figure \ref{fg:f1} shows the differences in magnitudes between the
infrared CIT and 2MASS photometric systems for the $J$, $H$, and
$K/K_S$ filters for the white dwarfs from Table 1. Note that the
number of stars in each panel is different (159 in $J$, 157 in $H$,
and $143$ in $K_S$) since some stars have not been formally detected
in one or more bands, and only lower limits are available. The size of
the error bars in Figure \ref{fg:f1} correspond to the combined
quadratic uncertainties of both data sets, $\sigma=(\sigma_{\rm
2MASS}^2+\sigma_{\rm CIT}^2)^{1/2}$. For both measurements to be
compatible, the error bar must touch the horizontal dashed line in
each panel of Figure \ref{fg:f1}, which represents the mean magnitude
difference between both data sets, as determined below.

We present in Table 2 a statistical comparison of both data sets for
all three bands. The first three lines correspond to the full
data set while the last three lines are restricted to 2MASS magnitudes
that satisfy the level~1 requirements. The second column indicates the
number of stars used for the comparison (to be included, the 2MASS
magnitude must have a measurement error). The third and fourth columns
represent respectively the mean and the standard deviation of the
magnitude differences for each band. These mean values thus
correspond to the zero point offsets between both photometric systems,
and we therefore adopt the following transformation based on the most
accurate subsample (level~1): $J_{\rm CIT} = J_{\rm 2MASS} - 0.0083$,
$H_{\rm CIT} = H_{\rm 2MASS} + 0.0094$ and $K_{\rm CIT} = K_{\rm
S~2MASS} + 0.0133$. We note that the offsets are typically five times
smaller than the average 2MASS uncertainties --- given in the fifth
column of Table 2, $\langle\sigma_{\rm 2MASS}\rangle$ --- and these
could as well be considered as zero for most practical purposes. We
also note that since the effective wavelength of the 2MASS $K_S$ filter
(2.169 $\mu$m) is slightly shorter than that of the CIT $K$ filter
(2.216 $\mu$m), the observed flux should be larger at $K_S$ than at
$K$, and a larger {\it positive} offset is thus expected for this
band, as is indeed observed in Table 2.

If the uncertainties of both data sets have been properly evaluated,
the average combined quadratic uncertainties, $\langle\sigma\rangle$
(last column of Table 2), should be at least as large as the standard
deviations of the magnitude differences (fourth column of Table 2).
This is certainly the case for the level~1 subsample, a result that
confirms the reliability of the 2MASS level~1 photometry. For the
complete sample, however, the $\langle\sigma\rangle$ values are
slightly below the standard deviations. If we assume that the CIT
photometric uncertainties have been properly estimated, which is
supported in BRL97 and BLR01 by the successful fits with white dwarf
models, the 2MASS uncertainties might be slightly underestimated in
the case of faint cool white dwarfs near the survey limit. 
Another way of interpreting these results is to note that in Figure
\ref{fg:f1}, the magnitudes are not compatible within the $1\sigma$
combined uncertainties for 34.6\%, 30.6\%, and 35.0\% of the stars in
the complete sample at the $J$, $H$ and $K$ bands,
respectively. These correspond to the objects whose error bars do not
cross the horizontal dashed lines.  This occurs
for level~1 and fainter objects as well. At a $3\sigma$ level, these
numbers drop to 0.6\%, 1.9\% and 4.2\%, respectively, which suggest
that there are infrequent but large discrepancies at $K_S$.

In Figure \ref{fg:f2}, we compare $(J-H,H-K/K_S)$ two-color
diagrams for various data sets. In the upper panels, we compare the
two-color diagrams for the 143 stars in common in both the CIT and
the 2MASS samples that have been detected by 2MASS in all three
bands. The 2MASS colors appear much more
scattered than the CIT colors, and this simply reflects the larger
uncertainties of the former data set. Indeed, if we restrict the
sample to the 49 objects that satisfy the level~1 requirements, the
scatter of the 2MASS diagram is greatly reduced, as shown in the
bottom panels of Figure \ref{fg:f2}. For this restricted sample, both
CIT and 2MASS data appear to have a similar scatter, which is a
confirmation of the comparable mean uncertainties. Since the 2MASS
photometry has been used to infer the presence of unresolved white
dwarf and low mass main sequence binaries, one needs to be cautious when
interpreting data sets that include objects below the level~1
requirements.

For instance, we indicated by open circles
in Figures \ref{fg:f1} and \ref{fg:f2} ten objects whose optical $BVRI$
and infrared $JHK$ photometry on the CIT system has been successfully
fitted with single white dwarf models by BRL97 and BLR01. They cover a
range in 2MASS $J$ magnitudes from 13.5 to 17. Our best fits for these
stars are displayed in Figure \ref{fg:f3}. The fitting technique used
here is described at length in BRL97. Briefly, the magnitudes on the
CIT system in Table 1 are first transformed onto the Johnson-Glass
system using the transformation equations given by
\citet{leggett92}. These magnitudes are then converted
into observed fluxes using the method described by
\citet{holberg06} for photon counting devices but using
the transmission functions taken from \citet{bessell90} for the $BVRI$
filters on the Johnson-Cousins photometric system, and from
\citet{bessell88} for the $JHK$ filters on the Johnson-Glass system.
The resulting energy distributions are then compared with those
predicted from our model atmosphere calculations, properly averaged
over the same filter bandpasses.  The hydrogen- and helium-rich model
atmospheres used in our analysis are similar to those described in
BLR01 and references therein, except that for the hydrogen-rich models
we are now making use of the more recent H$_2$-H$_2$ collision-induced
opacity calculations of
\citet{borysow01} and the Hummer-Mihalas occupation probability 
formalism for all species in the plasma. We find that the differences
in the fitted parameters are small compared to those derived by BLR01,
however.

The effective temperature $\Te$, the solid angle $\pi(R/D)^2$ (with
$R$ the radius of the star and $D$ its distance from Earth), and the
atmospheric composition (H- or He-rich) are obtained through a
$\chi^2$ minimization technique, where the $\chi ^2$ value is taken as
the sum over all bandpasses of the difference between observed and
predicted fluxes, properly weighted by observational
uncertainties. The trigonometric parallax measurement, when available,
is used to constrain the surface gravity through the mass-radius
relation for white dwarfs, otherwise a value of $\logg=8.0$ is
assumed. In Figure \ref{fg:f3}, the observed $BVRIJHK$ fluxes are
shown as error bars together with the monochromatic model fluxes (for
clarity, we do not show the average model fluxes at each bandpass).
The derived atmospheric parameters are given in each panel. As can be
seen, the energy distributions for all objects can be successfully
reproduced by assuming a single star model.

Also reproduced in Figure \ref{fg:f3} are the 2MASS magnitudes
converted into fluxes using the 2MASS zero points of
\citet{holberg06}. We note that for 9 of the 10 objects, at least one
of the fluxes at $J$, $H$, or $K_S$ is not
compatible with the predicted fluxes {\it within the $1\sigma$ 2MASS
uncertainties}. One exception is 0029$-032$, discussed later in \S~3,
for which the model spectrum matches the 2MASS photometry even better
than the CIT photometry. We thus conclude this section by stating that
while the 2MASS photometry is generally reliable, one should expect
occasional discrepancies. In particular, the detailed fits (not shown
here) to the energy distributions using the 2MASS photometry are of
good quality for most stars in our sample.

\section{WHITE DWARFS AND LOW MASS MAIN SEQUENCE BINARIES FROM 2MASS}

\subsection{The Wachter et al.~Analysis}

One of the most immediate applications to a large data set of white
dwarf NIR photometry such as 2MASS is to seek infrared excesses due to
cooler companions that are otherwise invisible in the
optical. \citet{wachter03} used a sample of 759 white dwarfs from the
catalog of \citet{ms99} and identified as many as 95 binary candidates
and 15 tentative binary candidates based on the analysis of a
$(J-H,H-K_S)$ two-color diagram built from 2MASS photometry. They
extracted $\JHKs$ magnitudes from the 2MASS second incremental data
release. Their binary candidates were selected from the color
criterion $(J-H)>0.4$, defined by the dashed horizontal lines in
our Figure \ref{fg:f2}, while their 15 tentative binary candidates
satisfy the criterion $0.2<(H-K_S)<0.5$ and $0.1<(J-H)<0.4$,
defined by the dotted rectangles in Figure
\ref{fg:f2}. In the following, we use the 2MASS final data release 
to recover more precise and slightly different observed $\JHKs$
magnitudes than those reported by Wachter et al.

Using the same color criteria to study the 2MASS sample of presumably
single cool white dwarfs presented in \S~2, we find in the upper-right
panel of Figure \ref{fg:f2} several binary and tentative binary
candidates in both regions defined by \citet{wachter03}. A comparison
with the CIT photometry, however, reveals that this result can be
readily explained in terms of the larger uncertainties of the 2MASS
photometry since both regions are located $1-2\sigma$ away from the region
occupied by single white dwarfs near the center of the figure. We find
that 3.5\% and 8.4\% of our sample observed by 2MASS contaminate the
binary candidate and tentative binary candidate regions,
respectively. By comparison, we find that at least 12.5\% of the white dwarfs in the
complete sample of 759 objects of Wachter et al.~are located in the
binary candidate region\footnote{The actual percentage may be larger
depending on how many faint objects with a partial detection are
removed from the sample.}. This indicates that the color criterion
defined to identify companions is certainly appropriate, but also that
the contamination from faint objects with large uncertainties near the
2MASS detection threshold may be significant. Furthermore, our large
contamination of the tentative binary candidate region suggests that this
criterion is not stringent enough, and that the corresponding
subsample identified by \citet[][Table 2]{wachter03} is mostly composed of single white
dwarfs.

These conclusions are supported by the fact that one of the objects
selected in the list of binary candidates (0102$+$210B) and four
objects in the list of tentative binary candidates (0029$-$032,
0518$+$333, 0816$+$387, and 1247$+$550)\footnote{We also found that
the 2MASS identification of 0145$-$174 by Wachter et al.~is erroneous;
the actual star is much fainter and not recovered in the 2MASS PSC.}
are all part of the single white dwarf sample described in
\S~2 and whose fits are displayed in Figure \ref{fg:f3}. As can be seen,
the CIT photometry for all objects is well reproduced with single star
model atmospheres.  For 0029$-$032, our fit is even better using the
2MASS photometry than the CIT data. For the other stars, the 2MASS
energy distributions appear flatter than those inferred from the CIT
photometry or the model spectra, a result that could be interpreted as
a flux excess in the $K$ band.

\subsection{The Wellhouse et al.~Analysis}

Using a similar approach but with slightly different criteria,
\citet{wellhouse05} sought companions to 51 magnetic white
dwarfs as candidates for potential pre-cataclysmic variables. They
proposed to split the $(J-H,H-K_S)$ two-color diagram into four
regions delimiting (I) single white dwarfs, (II) main sequence binary
candidates, (III) white dwarfs with very low mass companions, and (IV)
objects that may be contaminated by circumstellar material. These
representative regions are divided according to previous findings by
\citet{wachter03} as well as theoretical color simulations. While they 
did not find any convincing binary candidates (region II), Wellhouse
et al.~identified six objects with a possible very low mass companion
(region III) and four white dwarf candidates with an excess at $K_S$
(region IV), which they interpreted as a signature of undetected
planetary nebulae. This represents a total of 28.6\% of their sample
with formal uncertainties with a possible companion or a disk.

The four regions defined by \citet{wellhouse05} are reproduced here in
the $(J-H,H-K_S)$ two-color diagram shown in Figure \ref{fg:f4},
together with our common sample of CIT and 2MASS data composed of
presumably single white dwarfs. From this figure, we find that 21\%
of the white dwarfs in the 2MASS data set would be considered possible
candidates for a companion or a disk, while the CIT data show little
evidence for such infrared excesses. This strongly suggests that the
sample of magnetic white dwarfs studied by Wellhouse et al.~could be
entirely consistent with single stars. In addition, we note that among
the six objects located in region III of Figure~1 from Wellhouse et
al.~are some of the most intrinsically peculiar white
dwarfs\footnote{Also, 2201$-$228 in that sample is probably not
magnetic according to S.~Jordan (2005, private communication).}: LHS
2229 (1008$+$290) has been reported by \citet{schmidt99} and it has
the strongest C$_2$-like features ever observed, LP 790-29
(1036$-$204) is the strongest magnetic DQ known, and GD~229
(2010$+$310) shows strong unidentified absorption features in the
optical \citep[][Fig.~19]{wesemael93}. Therefore, region III seems to
be populated with some of the most peculiar white dwarfs for which
there is no reason to expect their NIR colors to overlap with those of
normal white dwarfs. Similarly, if we restrict our analysis to the
more accurate CIT data, there are three white dwarfs located in region
III of our Figure \ref{fg:f4}. Two of these identified in the figure
are also peculiar: G240-72 (1748$+$708) shows a deep yellow sag in the
4400-6300 \AA\ region \citep[][Fig.~19]{wesemael93}, and LP 701$-$29
(2251$-$070) is a heavily blanketed DZ star
\citep[][Fig.~11]{wesemael93}.

We also note that all four objects in region IV of
\citet[][Fig.~1]{wellhouse05} are very faint stars with 2MASS $K_S$
uncertainties in the range 0.16-0.27. As seen in our Figure
\ref{fg:f4}, we do expect single white dwarfs with large
uncertainties to populate this particular region as well. Hence the
location of the four objects identified by Wellhouse et al.~in this
particular region of the $(J-H,H-K_S)$ two-color diagram is most
naturally explained in terms of the low quality of the 2MASS data for
these objects rather than the presence of planetary nebulae. We thus
conclude that the identification of NIR excesses in the 2MASS PSC database
requires more conservative criteria allowing for larger
uncertainties in the photometric measurements below the level~1 requirements, or more
accurate methods such as that presented in the following section.

\subsection{The Synthetic Flux Method}

Another technique for identifying binary candidates is to compare observed
2MASS fluxes directly with those predicted from model atmospheres
\citep[see. e.g.,][]{holberg05,holberg06}. Effective temperatures and
surface gravities are first obtained using the spectroscopic method
developed by \citet{bergeron92} where high signal-to-noise
spectroscopic observations of the hydrogen Balmer lines are fitted
with synthetic models. The model flux is then normalized to the
observed $V$ magnitude to predict the observed fluxes at $J$, $H$, and
$K_S$ using the 2MASS filter passbands from
\citet{cohen03} and the zero points from \citet{holberg06}. Thus, only
objects with known atmospheric parameters and $V$ magnitudes can be
used with this method. In what follows, we rely on the fitting
technique and NLTE model atmospheres for DA stars described in
\citet{liebert05} and references therein.

To illustrate the method, we selected all DA stars from
\citet{wachter03} for which we had an optical spectrum and a published 
$V$ magnitude. In Table 3, we
present our sample that includes 42 binary candidates and 5 tentative
binary candidates from Tables 1 and 2 of
\citet{wachter03}, respectively\footnote{Note that the 2MASS
identification for 40~Eri~B (0413$+$077) by Wachter et al.~is
erroneous. With two objects within $2''$, they picked what is probably
the M dwarf 40~Eri~C instead of 40~Eri~B itself. Thus, while this is
still technically a WD$+$dM binary, both objects are barely resolved
in 2MASS and we do not include them in our sample.}. For each object,
we give the atmospheric parameters ($\Te$ and $\logg$), the published
$V$ magnitude, and the predicted and observed 2MASS magnitudes at $J$,
$H$, and $K_S$. In some cases, the optical spectrum was
significantly contaminated by the unresolved companion, and the
uncertainties on the derived parameters are correspondingly larger;
these are indicated by colons in Table 3.

For most objects, a significant NIR excess is observed, with the 2MASS
data being typically $\sim$2 magnitudes brighter than the values
predicted from the model fits. In Figure \ref{fg:f5}, we present
typical results for ten objects selected from Table 3. Here we show
the observed 2MASS fluxes together with the predicted monochromatic
fluxes calculated at the atmospheric parameters given in each
panel. For 0023$+$388, 0034$-$211, 0131$-$163, and 0145$-$257,
the companion can be unambiguously detected since the 2MASS fluxes are
about a factor of 10 to 100 larger than the predicted fluxes. For
0145$-$221, only a mild NIR excess is observed and this object has
indeed been identified as a WD$+$dL6/7 by \citet{farihi05} and
\citet{dobbie05}. Two of the tentative binary candidates, 0710$+$741
and 2257$+$162, do indeed show a significant excess consistent with a
very low mass companion. \citet{farihi05} have actually confirmed that
0710$+$741 is a WD$+$dM7. However, for 1434$+$289, 1639$+$153, and
2336$-$187, which are tentative binary candidates in
\citet{wachter03}, we do not observe any significant NIR excess and
these objects are thus consistent with being single white dwarfs.

With the exception of these last three objects, the infrared excesses
observed in Table 3 are consistent with unresolved low-mass main
sequence M dwarfs physically associated with the white dwarfs
\citep{farihi06}. Photometric observations of single M dwarfs by
\citet{leggett96} show that the $(J-V)$ color index is in the range
from $\sim2$ to 4, while single cool white dwarfs are expected to be
in the range from $-1$ to 1. This explains why the contribution of the
M dwarf can be dominant in the NIR but negligible in the optical. Many
of the 44 remaining binary candidates in Table 3 have been discussed
in the literature. For instance, \citet{farihi05} and \citet[][with
{\it HST}]{farihi06} observed 28 candidates from this list and were
able to resolve the red dwarf companion(s) for 17 objects. The NIR
excesses were also confirmed by \citet{farihi05,farihi06} using
$JHK$ photometric observations for the 11 remaining unresolved
objects. The presence of a companion for 9 additional objects in Table
3 has been discussed at various degrees in the literature, while for
the 7 remaining binary candidates (0812$+$478, 0915$+$201, 1037$+$512,
1108$+$325, 1339$+$346, 1610$+$383, and 2257$+$162), we confirm
through the synthetic flux method a strong NIR excess consistent with
the presence of low mass main-sequence companions.

We have seen that for a brown dwarf companion, the flux excess is not
as important as for M dwarfs. In the case of the dL6/7 dwarf companion
to 0145$-$221, the flux excess at $K_S$ is still significant at the
$12\sigma$ level, however, according to Table 3. There is only one
known example of a companion with a possible later spectral type, the
brown dwarf companion to 0137$-$349, discovered from radial velocity
measurements by
\citet{maxted06}, who also report a small excess at $K_S$ from 2MASS PSC
data. \citet{burleigh06} also present a near-IR spectrum that confirms
the slight $K$-band excess they attribute to a dL8 companion. We
analyzed the 2MASS photometry of this object with the method described
in this section, and assumed the effective temperature and surface
gravity from \citet{maxted06}. We were able to match very well the
predicted and observed 2MASS $J$ magnitude within the uncertainties,
and also identified a flux excess at $K_S$ at the $2.49\sigma$ level,
which is barely significant, but still consistent with the presense of a
disk or a companion. Therefore, the 2MASS survey is able to identify
hot brown dwarf companions, but it becomes more difficult to
confirm their presence for spectral types later than about
dL7.

We end this section by asserting that methods based on comparisons of
observed and predicted 2MASS fluxes (or magnitudes) represent an
efficient way of identifying unresolved white dwarf and low-mass main
sequence binaries down to late-type L dwarfs. Our analysis also
reveals, however, that $(J-H,H-K_S)$ two-color diagrams based on 2MASS
data should be interpreted with caution, and that regions expected to
contain unresolved binaries may be contaminated with single white
dwarfs, especially when data below the level~1 requirements are
considered.

\section{INFRARED PHOTOMETRY FROM SPITZER}

The {\it Spitzer Space Telescope} has been used to secure for the
first time IRAC 4.5 and $8~\mu$m photometric data for relatively
bright, nearby white dwarfs \citep[see, e.g.,][]{hansen06}. One of the
main interests of these surveys is to look for infrared flux excesses
due to the presence of circumstellar disks since it is expected that
the cool disk would dominate the MIR flux. It is however necessary as
a first step to evaluate the reliability of the {\it Spitzer} data set
and the ability of the model atmospheres to reproduce the MIR
fluxes. In such an effort, \citet{kilic06b} compared the {\it Spitzer}
4.5 and $8~\mu$m photometric data of 18 cool and bright white dwarfs
with the predictions of model atmospheres. They found that the four
hydrogen atmosphere white dwarfs with $\Te\lesssim6000$~K show a
slight flux depression at $8~\mu$m, while one peculiar object, the
so-called C$_2$H star LHS~1126, suffers from a significant flux
deficit at both 4.5 and $8~\mu$m. For the warmer objects, the model
fluxes seem to reproduce the {\it Spitzer} data perfectly.

In this section, we reanalyze 14 objects from the sample of
\citet{kilic06b} for which optical $BVRI$ photometry and infrared $JHK$ 
photometry on the CIT system are available (all of these are already part of
our cool white dwarf sample discussed in \S~2). In an approach similar
to that described in \S~2 (see Fig.~\ref{fg:f3}), we determine the
atmospheric parameters for each star by fitting simultaneously the
average fluxes for the nine photometric bands ($BVRI$, $JHK$/CIT,
and {\it Spitzer} 4.5 and $8~\mu$m). The synthetic fluxes in the MIR are
obtained by integrating our model grid over the {\it Spitzer} IRAC spectral
response curves while the observed fluxes are taken directly from
Table 1 of
\citet{kilic06b}.  In contrast with the technique used by Kilic et al.,
we do not normalize the fluxes at any particular band, but
consider instead the solid angle $\pi (R/D)^2$ a free parameter.
Since our $\chi^2$ value is taken as the sum over all bands of
the difference between observed and model fluxes, properly weighted by
the corresponding observational errors, our approach has the advantage
of allowing for the full photometric uncertainties in the fitting
procedure. Furthermore, instead of assuming $\logg=8.0$ for
all objects, we constrain the $\logg$ value from the trigonometric
parallax measurements, as described above.

In Figure \ref{fg:f6}, we present our best fits on a logarithmic scale
to the observed $BVRI$, $JHK$ (CIT), and {\it Spitzer} photometry with the
model average fluxes described above. We also plot the monochromatic
fluxes for clarity; the case of LHS~1126 is discussed separately
below. Another peculiar object, G240-72 (1748+708) already discussed
near the end of \S~3.2, shows a deep unidentified absorption in the optical (a yellow
sag) and no satisfactory fit can be achieved for this star and it is
thus left out of our analysis. For Ross 627 (1121$+$216), the $8~\mu$m
flux is not shown in Figure \ref{fg:f6} since \citet{kilic06b} provides
only an upper limit due to a possible contamination from a nearby
star.  Our final sample thus includes 12 stars with 23 {\it Spitzer} 4.5 and
$8~\mu$m flux measurements. For all cases shown in Figure
\ref{fg:f6}, the {\it Spitzer} fluxes are well reproduced by the synthetic
models. To further strengthen this conclusion, we plot in Figure
\ref{fg:f7} the ratio of the observed to model fluxes at
4.5 and $8~\mu$m as a function of the derived effective temperature
for the 12 objects. The figure confirms the agreement between the
observed {\it Spitzer} and model fluxes at all temperatures. In particular,
we do not observe any significant flux deficit at low effective
temperatures as suggested by \citet{kilic06b}. There are only 2
observations out of 23 for which the flux deficit is significant at
the $1\sigma$ level, and both are in the $8~\mu$m band. It thus
seems premature to conclude from these results that there is any
discrepancy between the observations and the predictions of
model atmospheres with pure hydrogen compositions. 

We mention in this context that the second coolest object in Figure
\ref{fg:f7} is the DA star BPM 4729 (0752$-$676) for which we obtain a
perfect fit. This star has been studied extensively by BLR01, and more
recently by \citet{kowalski06} using improved L$\alpha$ profiles that
include broadening by molecular hydrogen, and both atmospheric
parameter determinations agree at the $1\sigma$ level under the
assumption of pure hydrogen compositions. Hence for this well studied
normal cool DA star, independent model atmospheres yield consistent
atmospheric parameters that both match the observational data.  In
contrast, the two objects --- LHS 1038 (0009$+$501) and G99-47
(0553$+$053) --- with the small $8~\mu$m flux discrepancies (bottom
panel of Fig.~\ref{fg:f7}) are magnetic white dwarfs. Both objects
show $1\sigma$ discrepancies at $J$ and also at $B$ for G99-47. While
this suggests that the inclusion of a magnetic field in the model
atmosphere calculations could improve the fit, we believe that the
discrepancies observed here are only barely significant and not
systematic enough to make formal conclusions. Therefore, we argue that
the results presented in this section demonstrate the reliability of
both the {\it Spitzer} IRAC photometry and our model atmosphere grid
up to $8~\mu$m for studying cool white dwarfs. The consistency between
models and data is critical for surveys seeking MIR infrared excesses
from circumstellar disks. Our results indicate that the comparison of
{\it Spitzer} fluxes with theoretical predictions could identify such
MIR excesses with relatively high precision.

In an attempt to identify the nature of the discrepancy between our
conclusions and those reached by \citet{kilic06b}, we have performed
the same analysis as above but with the 2MASS $JHK_s$ magnitudes used
by Kilic et al.~(instead of the CIT magnitudes used in this
analysis). We have also tried to normalize our solutions at $V$, as
done by Kilic et al. In all of our experiments, the results differ
only slightly from those reported here, and our main conclusions thus
remain the same. We are therefore unable to explain the differences
between both studies.  We can only emphasize that the analysis of {\it
Spitzer} photometric data appears to be sensitive to the details of
the fitting procedure.

Another white dwarf analyzed by \citet{kilic06b} is LHS~1126
(0038$-$226) whose energy distribution is characterized by a strong
infrared flux deficiency at $JHK$ interpreted by \citet{bergeron94} in
terms of collision-induced absorption (CIA) by molecular hydrogen due
to collisions with helium in a mixed hydrogen and helium atmosphere
with $N({\rm H})/N({\rm He})\sim 0.01$. We do confirm here the results
shown in Figure 4 of \citet{kilic06b} where the {\it Spitzer} fluxes
are significantly depressed with respect to the predictions of model
atmospheres with mixed compositions. The main reason for this
discrepancy is that the CIA opacity predicts a {\it maximum}
absorption near the H$_2$ fundamental vibration frequency at
$\sim2.4\mu$m, while the {\it Spitzer} fluxes are more consistent with
a featureless energy distribution from 1 to $8~\mu$m. This problem is
surprisingly similar to that encountered in the so-called ultra-cool
white dwarfs, and in particular in the case of LHS~3250 for which the
H$_2$-H$_2$ and H$_2$-He CIA opacities predict absorption bands that
are simply not observed in spectroscopy \citep{bergeron02}. These
results may indicate that the collision-induced opacity calculations
need to be improved at the high densities encountered in cool white
dwarf atmospheres.

\section{CANDIDATE WHITE DWARFS WITH CIRCUMSTELLAR DISKS}

The synthetic flux method based on a comparison of predicted and
observed 2MASS fluxes (or magnitudes) was shown to be an efficient
technique for detecting NIR excesses from unresolved companions
(\S~3). However, the NIR excess in the $\JHKs$ bands expected from
cool circumstellar disks or planets surrounding white dwarf stars can
be extremely small if the flux is dominated by the white dwarf in this
particular wavelength range. In this section, we use the results of
the ongoing spectroscopic survey of \citet{gianninas06} together with
the 2MASS PSC to search for massive disks resulting from the merger of
two white dwarfs, as predicted by \citet{livio05}. In addition to the
synthetic flux method described above, we also compare the observed
and predicted $(J-H)$ and $(J-K_S)$ color indices since this method
has the advantage of being independent of the normalization at $V$,
which allows us to consider also objects with no published $V$
magnitudes. Since circumstellar disks are expected to be much brighter
at $K_S$ than in the other bands, we expect their color indices to be
very different from those of single white dwarfs, and such objects
should easily stand out in our analysis.

As discussed in the Introduction, white dwarfs resulting from mergers
are expected to be found in the high-mass tail of the mass
distribution.  We thus selected all DA stars from the survey of
\citet{gianninas06} with spectroscopic masses above 0.8 \msun\ that
were formally detected by 2MASS in at least two bands (usually the $J$
and $H$ bands), for a total of 57 objects. In Table 4, we provide the
effective temperature, the spectroscopic mass, the $V$ magnitude (when
available), as well as the predicted and observed 2MASS $\JHKs$
magnitudes for each object in our sample. The atmospheric parameters
($\Te$ and $\logg$) are obtained from fits to the Balmer lines using
the NLTE model grid described in \S~3, and the $\logg$ values are
converted into mass using the evolutionary models of \citet{wood95}
with carbon-core compositions and thick hydrogen layers. The predicted
fluxes are obtained from the synthetic flux method and are thus only
given for objects with measured $V$ magnitudes.

Five white dwarfs in Table 4 (0429$+$176, 0950$+$139,
1058$-$129, 1120$+$439, and 1711$+$668) show a large NIR flux excess
that is not attributable to a circumstellar disk. The predicted
spectra for these stars are shown in Figure \ref{fg:f8} together with
the observed 2MASS fluxes. We discuss each object in turn.

\noindent{\it HZ 9 (0429$+$176)} -- This object is a WD$+$dM binary
\citep{lanning81} in common with the sample discussed in \S~3.

\noindent{\it PG 0950$+$139} -- This star is in common with the sample
discussed in \S~3. The white dwarf is surrounded by a planetary
nebulae \citep{ellis84} and its optical spectrum exhibits emission
lines \citep{liebert89}. According to \citet{fulbright93}, the
low-density gas emission and the infrared excess are best explained by
the presence of a low-mass companion.

\noindent{\it PG 1058$-$129, PG 1120$+$439} -- These two
objects show a mild and unexplained infrared excess. In both cases,
the only $V$ magnitudes available are multichannel data from the
Palomar-Green survey \citep{PG}.  Since the observed energy slopes
measured by color indices are in perfect agreement with those
predicted by the models (see below), it is very likely that the $V$
magnitudes for these stars are simply erroneous. We note that
\citet{green00} also determined a 1-sigma significant excess at $J$
for PG 1120$+$439.

\noindent{\it RE J1711+664 (1711$+$668)} -- This white dwarf is a
barely resolved visual pair \citep{finley97}. The predicted NIR flux
from this white dwarf is too low to be detected by 2MASS. Thus only
the dM star $\sim 2''$ away from the white dwarf is detected in the
PSC.

We exclude from our analysis the three objects with known companions,
but we keep PG 1058$-$129 and PG 1120$+$439.  

We compare in Figure \ref{fg:f9} the observed and predicted $(J-H)$
and $(H-K_S)$ color indices as a function of $H$ and $K_S$,
respectively, for the remaining 54 white dwarfs in our sample. An
examination of these results indicate that all stars are consistent
with the predicted white dwarf colors within $3\sigma$ uncertainties,
both above and below the level 1 requirements. Two glaring exceptions
are G1-7 (0033+016) and CBS 413 (1554+322), which are among the
faintest objects in the bottom panel of Figure \ref{fg:f9} (labeled
1 and 2, respectively). For G1-7, however, the color
indices derived from the CIT photometry given in Table 1 are in
excellent agreement with those predicted by the models. Also, CBS 413
has not been detected at $H$ but it is unexpectedly bright at $K_S$!
Since this object has no published $V$ magnitude, it is not clear
whether the $J$ detection is indeed from the white dwarf, and thus
whether the color excess at $K_S$ is even real. Therefore, we conclude
from the results shown in Figure
\ref{fg:f9} that there is no strong evidence for $H$ or $K_S$ excesses
in this sample of massive white dwarfs, and for the presence of
massive circumstellar disks around them.

For comparison, we also reproduce in Figure \ref{fg:f9} the location
of three white dwarfs with previously identified circumstellar disks:
G29-38 (2326+049), GD 362 (1729+371) and GD 56 (0408$-$041). The
atmospheric parameters for all three stars have been determined using
our own spectroscopic observations, and the predicted 2MASS color
indices have been estimated from the same method as above.  For the
metal-rich DAZ star GD~362, we use the more accurate atmospheric
parameters of \citet{gianninas04} who took into account the presence
of heavy elements in their model atmosphere calculations. Only GD~362
in our sample is a massive white dwarf with $M=1.24$ \msun, while we
obtain $M=0.70$ and 0.60 \msun\ for G29-38 and GD~56, respectively.
The disk around G29-38 was the first discovered and studied
extensively in the MIR \citep{reach05}. The object is clearly
identifiable in Figure \ref{fg:f9} with $(J-K_S)_{\rm
2MASS}-(J-K_S)_{\rm pred}=0.52\pm 0.04$, a $\sim 12\sigma$ result. The
second object, GD 362, is a massive DAZ star for which
\citet{becklin05} reported the discovery of an important flux excess
at $L'$ ($3.76~\mu$m) and $N'$ ($11.3~\mu$m). \citet{kilic05} obtained
a near infrared spectrum in the $0.8-2.5~\mu$m range but found only a
mild flux excess at $K$. Both studies concluded that the presence of a
dust disk could account for the observations. Given that GD 362 is
particularly faint ($V=16.3$), only lower limits at $H$ and $K_S$ are
available in the 2MASS PSC. Instead, we use in Figure \ref{fg:f9} the
$\JHKs$ magnitudes measured by \citet{becklin05}.  With these
measurements, GD 362 exhibits a color excess of $(J-K_S)_{\rm
obs}-(J-K_S)_{\rm pred}=0.22\pm 0.04$, a $5\sigma$ result.
Unfortunately, this photometric accuracy is only achieved in the 2MASS
sample for $J$ brighter than $\sim 14.1$, and a color excess of the
magnitude found in GD 362 cannot be easily uncovered in the majority
of white dwarfs detected by 2MASS. For GD 56, \citet{kilic06a}
reported a NIR excess in both the 2MASS data and in their own infrared
spectroscopic observations.  Unlike the two previous objects, GD 56
lacks the MIR observations that could confirm the presence of a
disk. We recovered the 2MASS magnitudes from the PSC and determined a
color excess of $(J-K_S)_{\rm 2MASS}-(J-K_S)_{\rm pred}=0.54\pm 0.19$,
a $2.9\sigma$ result, barely significant according to our $3\sigma$
criterion.

From the analysis of the three known white dwarfs with circumstellar
disks, we conclude that the infrared excess from similar disks around
white dwarfs would be significant only for bright level~1 2MASS
objects. We argue that while the 2MASS PSC is indeed able to suggest
the presence of a disk for fainter stars like GD~56, MIR photometric
observations or more accurate NIR data would be required to
unambiguously identify circumstellar disks such as those discussed
here. Furthermore, according to \citet{livio05}, a circumstellar disk
resulting from the merger of two white dwarfs would presumably have a
much larger mass and radius in comparison with the disks currently
known. Hence the expected NIR excess should also be large. Obviously,
such large infrared excesses have not been detected in our 2MASS
sample, and we conclude that massive circumstellar disks are uncommon
around massive white dwarfs, in agreement with the conclusions reached
by \citet{hansen06} based on {\it Spitzer} data. While our results
constrain the scenario proposed by \citet{livio05}, the fraction of
massive degenerates in our sample that are the product of white dwarf
mergers is totally unknown. For instance, \citet{dobbie06} suggested
that GD 50 (0346$-$011) is associated with the star formation event
that created the Pleiades, and this massive white dwarf is most likely
a former member of this cluster. Hence the authors find no need to
invoke a double white dwarf merger scenario to account for its
existence. Thus, massive circumstellar disks may not be expected in
all cases studied here.

\section{CONCLUSION}

In order to estimate the reliability of the 2MASS photometry for white
dwarf stars, we defined a sample of 160 cool degenerates with $JHK$
magnitudes on the CIT photometric system taken from BRL97 and BLR01,
and compared these values with those obtained from the 2MASS PSC. Our
statistical analysis indicates that, on average, both data sets are
consistent within the uncertainties, and thus that the 2MASS
photometric data is appropriate for the study of white dwarf
stars. The 2MASS data should still be interpreted with caution,
however, especially for stars near the detection threshold, as
significant discrepancies are to be expected.

We also concluded that the search for white dwarf and main-sequence
star binaries based on 2MASS two-color diagrams is greatly limited by
the 2MASS uncertainties when data below the level~1 requirements are
considered. We demonstrated that some color regions identified by
\citet{wachter03} and \citet{wellhouse05} to search for binary
candidates are highly contaminated by single stars. We analyzed 47
binary candidates taken from the sample of \citet{wachter03} using the
synthetic flux method and showed that this technique is a much more
efficient tool for confirming binary candidates.  We have also shown
that the observed MIR photometry from the {\it Spitzer Space
Telescope} agree very well with our model fluxes, a result that
confirms the reliability of both the {\it Spitzer} photometry and our
model atmosphere calculations up to $8~\mu$m.

Finally, we searched for massive and large circumstellar disks, such
as those predicted by \citet{livio05}, around 57 massive white dwarfs
($M > 0.8$ \msun). We showed that these systems would be clearly
distinguishable from single stars in the 2MASS PSC, but such systems
have not yet been identified in our analysis. Hence, high-mass
circumstellar disks resulting from the merger of two white dwarfs must
be uncommon around massive white dwarfs. We also showed that low-mass
circumstellar disks such as those associated with G29-38, GD~362 and
GD~56 are only barely identifiable except perhaps for the brightest
level~1 white dwarfs in the 2MASS PSC.

We would like to thank A.~Gianninas for a careful reading of our
manuscript and for sharing the results of his ongoing spectroscopic
survey. This work was supported in part by the NSERC
Canada. P. Bergeron is a Cottrell Scholar of Research
Corporation. This publication makes use of data products from the Two
Micron All Sky Survey, which is a joint project of the University of
Massachusetts and the Infrared Processing and Analysis
Center/California Institute of Technology, funded by the National
Aeronautics and Space Administration and the National Science
Foundation. This work is based in part on observations made with the
{\it Spitzer Space Telescope}, which is operated by the Jet Propulsion
Laboratory, California Institute of Technology under a contract with
NASA.

\clearpage
\clearpage
\begin{deluxetable}{llcccccccc}
\tabletypesize{\scriptsize}
\tablecolumns{10}
\tablewidth{0pt}
\tablecaption{Sample of Cool White Dwarfs with Near-Infrared Photometry}
\tablehead{
\colhead{WD} &
\colhead{Name} &
\colhead{$J_{\rm CIT}$} &
\colhead{$H_{\rm CIT}$} &
\colhead{$K_{\rm CIT}$} &
\colhead{$J_{\rm 2MASS}$ ($\sigma_J$)} &
\colhead{$H_{\rm 2MASS}$ ($\sigma_H$)} &
\colhead{$K_{S\,{\rm 2MASS}}$ ($\sigma_K$)}}
\startdata
 0000$-$345  &LHS 1008      
 &14.17&14.02&13.87&14.117 (0.024) &14.024 (0.038)& 13.919 (0.063)\\     
 0007$+$308  &LHS 1028      
 &16.43&16.34&16.33&16.449 (0.128) &16.193 (0.224)& 16.614 (null)\\      
 0009$+$501  &LHS 1038      
 &13.41&13.26&13.21&13.490 (0.022) &13.249 (0.026)& 13.191 (0.030)\\     
 0011$+$000  &G31-35      
 &15.21&15.13&15.12&15.148 (0.039) &15.214 (0.094)& 15.101 (0.139)\\     
 0011$-$134  &LHS 1044      
 &14.85&14.62&14.52&14.813 (0.036) &14.549 (0.057)& 14.628 (0.082)\\     
 0029$-$032  &LHS 1093      
 &15.56&15.37&15.35&15.635 (0.050) &15.380 (0.091)& 15.166 (0.147)\\     
 0033$+$016  &G1-7        
 &15.63&15.60&15.63&15.650 (0.057) &15.522 (0.090)& 16.119 (0.303)\\     
 0038$-$226  &LHS 1126      
 &13.32&13.47&13.71&13.342 (0.028) &13.483 (0.033)& 13.738 (0.044)\\     
 0038$+$555  &G218-8      
 &13.97:&14.08:&14.13:&14.066 (0.036) &13.981 (null)& 13.967 (null)\\    
 0046$+$051  &vMa 2         
 &11.69&11.61&11.52&11.688 (0.022) &11.572 (0.024)& 11.498 (0.025)\\     
 0048$-$207  &LHS 1158      
 &15.76&15.38&15.38&15.748 (0.060) &15.378 (0.090)& 15.216 (0.125)\\     
 0101$+$048  &G1-45       
 &13.51&13.39&13.38&13.504 (0.024) &13.396 (0.032)& 13.418 (0.034)\\     
 0102$+$210A &LHS 5023      
 &16.56:&16.32:&16.21:&16.518 (0.097) &16.504 (0.198)& 15.548 (null)\\   
 0102$+$210B &LHS 5024      
 &16.45:&16.12:&16.00:&16.734 (0.110) &16.267 (0.164)& 15.589 (0.188)\\  
 0112$-$018  &LHS 1219      
 &16.29&16.00&15.97&16.288 (0.089) &15.763 (0.136)& 15.974 (0.261)\\     
 0115$+$159  &LHS 1227      
 &13.72&13.72&13.74&13.727 (0.025) &13.680 (0.022)& 13.726 (0.044)\\     
 0117$-$145  &LHS 1233      
 &15.47&15.17&15.08&15.563 (0.056) &15.131 (0.079)& 15.192 (0.161)\\     
 0121$+$401  &G133-8      
 &15.64&15.43&15.41&15.858 (0.078) &15.507 (0.151)& 15.279 (0.170)\\     
 0123$-$262  &LHS 1247      
 &14.50&14.36&14.38&14.435 (0.029) &14.313 (0.044)& 14.331 (0.072)\\     
 0126$+$101  &G2-40       
 &14.05&13.92&13.94&14.032 (0.024) &13.952 (0.038)& 13.964 (0.053)\\     
 0135$-$052  &L870-2      
 &12.12&11.94&11.92&12.114 (0.024) &11.954 (0.022)& 11.969 (0.023)\\     
 0142$+$312  &G72-31      
 &14.38&14.33&14.38&14.425 (0.029) &14.320 (0.048)& 14.429 (0.065)\\     
 0208$+$396  &G74-7       
 &13.80&13.65&13.63&13.832 (0.024) &13.670 (0.034)& 13.595 (0.038)\\     
 0222$+$648  &LHS 1405      
 &16.35&16.06&15.98&16.357 (0.100) &15.572 (0.125)& 15.729 (0.212)\\     
 0230$-$144  &LHS 1415      
 &14.43&14.17&14.11&14.489 (0.030) &14.261 (0.048)& 14.161 (0.068)\\     
 0243$-$026  &LHS 1442      
 &14.71&14.49&14.51&14.679 (0.035) &14.589 (0.044)& 14.477 (0.091)\\     
 0245$+$541  &LHS 1446      
 &13.89&13.66&13.60&13.870 (0.024) &13.545 (0.040)& 13.469 (0.039)\\     
 0322$-$019  &G77-50      
 &14.63&14.37&14.28&14.761 (0.042) &14.439 (0.052)& 14.378 (0.084)\\     
 0326$-$273  &L587-77A    
 &13.27&13.12&13.08&13.216 (0.103) &13.109 (0.090)& 13.101 (0.121)\\     
 0341$+$182  &Wolf 219      
 &14.56&14.35&14.40&14.590 (0.031) &14.350 (0.049)& 14.230 (0.060)\\     
 0357$+$081  &LHS 1617      
 &14.59&14.33&14.26&14.562 (0.038) &14.343 (0.056)& 14.122 (0.057)\\     
 0407$+$197  &LHS 1636      
 &16.26&16.03&15.82&16.130 (0.084) &15.957 (0.161)& 15.178 (null)\\      
 0423$+$044  &LHS 1670      
 &15.50&15.29&15.26&15.474 (0.068) &15.182 (0.075)& 15.168 (0.150)\\     
 0423$+$120  &G83-10      
 &14.52&14.34&14.27&14.485 (0.034) &14.347 (0.042)& 14.249 (0.065)\\     
 0433$+$270  &G39-27      
 &14.61&14.32&14.22&14.598 (0.038) &14.232 (0.058)& 14.136 (0.069)\\     
 0435$-$088  &L879-14     
 &13.00&12.85&12.79&13.006 (0.030) &12.906 (0.032)& 12.763 (0.035)\\     
 0437$+$093  &LHS 1693      
 &16.03&15.80&15.81&15.944 (0.075) &15.583 (0.102)& 15.583 (0.158)\\     
 0440$+$510  &G175-46     
 &15.60&15.50&15.53&15.576 (0.051) &15.504 (0.112)& 15.548 (0.141)\\     
 0503$-$174  &LHS 1734      
 &14.55&14.33&14.23&14.739 (0.035) &14.408 (0.047)& 14.397 (0.086)\\     
 0511$+$079  &G84-41      
 &15.01&14.79&14.71&15.107 (0.047) &14.924 (0.064)& 14.863 (0.082)\\     
 0518$+$333  &G86-B1B     
 &15.52&15.33&15.34&15.374 (0.085) &15.122 (0.207)& 14.806 (0.157)\\     
 0548$-$001  &G99-37      
 &13.73&13.63&13.63&13.730 (0.029) &13.675 (0.026)& 13.705 (0.043)\\     
 0551$+$468  &LHS 1801      
 &15.84&15.55&15.53&15.712 (0.057) &15.461 (0.078)& 15.511 (0.154)\\     
 0552$-$041  &LP 658-2    
 &13.02&12.90&12.82&13.047 (0.027) &12.860 (0.027)& 12.777 (0.026)\\     
 0553$+$053  &G99-47      
 &12.96&12.77&12.66&12.930 (0.022) &12.720 (0.025)& 12.653 (0.024)\\     
 0618$+$067  &LHS 1838      
 &15.29&15.05&15.00&15.377 (0.062) &15.017 (0.071)& 14.957 (0.139)\\     
 0644$+$025  &G108-26     
 &15.00&14.85&14.93&14.868 (0.045) &14.757 (0.069)& 14.576 (0.103)\\     
 0648$+$641  &LP 58-53    
 &15.46&15.19&15.12&15.533 (0.061) &15.412 (0.098)& 15.331 (0.165)\\     
 0654$+$027  &G108-42     
 &15.98&15.98&15.98&16.086 (0.088) &15.824 (0.133)& 15.399 (0.188)\\     
 0657$+$320  &LHS 1889      
 &14.99&14.77&14.69&15.030 (0.039) &14.674 (0.050)& 14.665 (0.082)\\     
 0659$-$064  &LHS 1892      
 &14.58&14.29&14.24&14.538 (0.028) &14.218 (0.051)& 14.355 (0.074)\\     
 0704$-$508  &ESO 207-124 
 &16.24::&16.11::& null&16.08 (0.062) &16.093 (0.131)& 15.688 (0.196)\\  
 0706$+$377  &G87-29      
 &15.00&14.88&14.82&15.064 (0.053) &14.783 (0.052)& 14.834 (0.079)\\     
 0738$-$172  &L745-46A    
 &12.65&12.61&12.52&12.653 (0.022) &12.611 (0.026)& 12.583 (0.036)\\     
 0747$+$073A &LHS 240       
 &14.96&14.73&14.72&14.996 (0.039) &14.719 (0.067)& 14.634 (0.099)\\     
 0747$+$073B &LHS 239       
 &15.05&14.90&14.86&15.031 (0.037) &14.898 (0.080)& 14.746 (0.107)\\     
 0751$+$578  &G193-78     
 &14.94&14.94&14.96&14.966 (0.038) &14.965 (0.063)& 14.966 (0.121)\\     
 0752$-$676  &BPM 4729      
 &12.79&12.52&12.43&12.726 (0.023) &12.476 (0.026)& 12.362 (0.024)\\     
 0752$+$365  &G90-28      
 &15.50&15.35&15.35&15.583 (0.064) &15.444 (0.131)& 15.877 (0.346)\\     
 0802$+$386  &LP 257-28   
 &15.60&15.58&15.64&15.376 (null) &15.663 (0.150)& 15.046 (null)\\       
 0802$+$387  &LHS 1980      
 &15.26&15.02&14.96&15.336 (0.047) &15.193 (0.079)& 14.899 (0.091)\\     
 0806$-$661  &L97-3       
 &13.79&13.85&13.92&13.704 (0.023) &13.739 (0.025)& 13.781 (0.043)\\     
 0813$+$217  &G40-15      
 &16.03&15.74&15.65&15.944 (0.068) &15.842 (0.142)& 15.958 (0.240)\\     
 0816$+$387  &G111-71     
 &15.87&15.72&15.73&16.070 (0.103) &15.833 (0.194)& 15.583 (0.220)\\     
 0827$+$328  &LHS 2022      
 &15.01&14.85&14.84&14.985 (0.044) &14.964 (0.076)& 14.865 (0.121)\\     
 0839$-$327  &L532-81     
 &11.59&11.55&11.55&11.578 (0.030) &11.539 (0.033)& 11.547 (0.029)\\     
 0856$+$331  &G47-18      
 &15.12&15.09&15.11&15.172 (0.041) &15.156 (0.083)& 15.312 (0.163)\\     
 0912$+$536  &G195-19     
 &13.22&13.15&13.09&13.308 (0.025) &13.211 (0.026)& 13.133 (0.030)\\     
 0913$+$442  &G116-16     
 &14.96&14.84&14.87&14.955 (0.050) &14.861 (0.081)& 14.906 (0.155)\\     
 0930$+$294  &G117-25     
 &15.51&15.40&15.45&15.588 (0.066) &15.399 (0.106)& 15.284 (0.150)\\     
 0941$-$068  &G161-68     
 &15.35&15.06&15.08&15.373 (0.042) &15.019 (0.069)& 14.990 (0.133)\\     
 0946$+$534  &G195-42     
 &14.90&14.87&14.88&14.913 (0.049) &14.888 (0.072)& 14.916 (0.118)\\     
 0955$+$247  &G49-33      
 &14.66&14.59&14.65&14.654 (0.034) &14.659 (0.069)& 14.661 (0.076)\\     
 1012$+$083  &G43-38      
 &15.20&14.99&14.94&15.246 (0.063) &15.132 (0.110)& 14.955 (0.142)\\     
 1019$+$637  &LP 62-147   
 &13.83&13.63&13.65&13.874 (0.029) &13.733 (0.047)& 13.692 (0.049)\\     
 1026$+$117  &LHS 2273      
 &15.93&15.83&15.68&15.902 (0.095) &15.552 (0.127)& 15.298 (null)\\      
 1039$+$145  &G44-32      
 &15.93&15.86&15.78&15.823 (0.069) &15.749 (0.156)& 15.681 (0.205)\\     
 1055$-$072  &LHS 2333      
 &13.81&13.71&13.69&13.770 (0.029) &13.680 (0.032)& 13.485 (0.038)\\     
 1108$+$207  &LHS 2364      
 &15.91&15.69&15.62&15.978 (0.074) &15.532 (0.107)& 15.565 (0.161)\\     
 1114$+$067  &G45-45      
 &15.82&15.57&15.54&15.701 (0.072) &15.599 (0.124)& 15.518 (0.257)\\     
 1115$-$029  &LHS 2392      
 &15.23&15.27:&15.29:&15.304 (0.051) &15.246 (0.084)& 15.734 (0.241)\\   
 1121$+$216  &Ross 627      
 &13.58&13.40&13.40&13.574 (0.024) &13.420 (0.026)& 13.399 (0.034)\\     
 1124$-$296  &ESO 439-80  
 &14.90&14.88&14.80&14.782 (0.034) &14.710 (0.044)& 14.602 (0.091)\\     
 1142$-$645  &LHS 43        
 &11.19&11.12&11.09&11.188 (0.024) &11.130 (0.025)& 11.104 (0.026)\\     
 1146$-$291  &ESO 440-146 
 &16.17&15.89:&15.71:&16.037 (0.092) &15.737 (0.172)& 16.462 (null)\\    
 1147$+$255  &LP 375-51   
 &15.53&15.49&15.51&15.590 (0.048) &15.568 (0.101)& 15.693 (0.183)\\     
 1153$+$135  &LHS 2478      
 &16.10&15.83&15.74:&15.889 (0.062) &15.503 (0.095)& 15.575 (0.149)\\    
 1154$+$186  &LP 434-97   
 &15.15&15.03&15.02&15.098 (0.042) &15.220 (0.099)& 15.087 (0.145)\\     
 1208$+$576  &LHS 2522      
 &14.64&14.39&14.32&14.679 (0.034) &14.362 (0.052)& 14.458 (0.095)\\     
 1236$-$495  &LTT 4816      
 &13.92&13.90&13.98&13.806 (0.024) &13.815 (0.036)& 13.907 (0.062)\\     
 1239$+$454  &LHS 2596      
 &15.47&15.30&15.30&15.599 (0.062) &15.197 (0.101)& 15.727 (null)\\      
 1244$+$149  &G61-17      
 &15.84&15.86&15.84&15.802 (0.067) &15.627 (0.136)& 15.721 (0.217)\\     
 1247$+$550  &LP 131-66   
 &15.72&15.67&15.63&15.795 (0.067) &15.659 (0.131)& 15.396 (0.212)\\     
 1257$+$037  &LHS 2661      
 &14.56&14.33&14.25&14.655 (0.040) &14.316 (0.050)& 14.220 (0.089)\\     
 1257$+$278  &G149-28     
 &14.99&14.91&14.91:&15.132 (0.046) &14.977 (0.076)& 14.986 (0.089)\\    
 1300$+$263  &LHS 2673      
 &16.89&16.71&16.70&16.801 (0.142) &16.399 (0.214)& 16.436 (null)\\      
 1310$-$472  &ER 8          
 &15.21&15.11&15.03&15.135 (0.045) &15.045 (0.080)& 14.735 (0.123)\\     
 1313$-$198  &LHS 2710      
 &15.87&15.70&15.56&15.875 (0.082) &15.612 (0.097)& 15.550 (0.190)\\     
 1325$+$581  &G199-71     
 &15.82&15.68&15.65&15.945 (0.092) &15.700 (0.147)& 15.716 (0.251)\\     
 1328$+$307  &G165-7      
 &15.50&15.36&15.34&15.402 (0.044) &15.282 (0.087)& 15.413 (0.135)\\     
 1330$+$015  &G62-46      
 &16.38&16.25&16.17:&16.396 (0.119) &16.298 (0.206)& 15.802 (null)\\     
 1334$+$039  &Wolf 489      
 &13.06&12.80&12.70&13.064 (0.024) &12.819 (0.026)& 12.690 (0.021)\\     
 1344$+$106  &LHS 2800      
 &14.38&14.20&14.19&14.407 (0.038) &14.139 (0.053)& 14.235 (0.080)\\     
 1345$+$238  &LP 380-5    
 &13.92&13.67&13.59&13.921 (0.027) &13.669 (0.036)& 13.621 (0.040)\\     
 1346$+$121  &LHS 2808      
 &16.52&16.43&16.32&16.463 (0.118) &16.193 (0.242)& 15.810 (null)\\      
 1418$-$088  &G124-26     
 &14.81&14.69&14.69&14.764 (0.037) &14.731 (0.057)& 14.756 (0.103)\\     
 1444$-$174  &LHS 378       
 &14.94&14.79&14.68&14.948 (0.029) &14.640 (0.047)& 14.724 (0.108)\\     
 1455$+$298  &LHS 3007      
 &14.86&14.73&14.72&14.972 (0.047) &14.606 (0.075)& 14.739 (0.128)\\     
 1503$-$070  &GD 175        
 &15.07&14.93&14.91&15.079 (0.052) &14.988 (0.100)& 14.847 (0.104)\\     
 1602$+$010  &LHS 3151      
 &16.08&15.86&15.67&16.078 (0.081) &15.969 (0.173)& 15.526 (0.176)\\     
 1606$+$422  &Case 2        
 &13.92&13.92&14.01&13.984 (0.025) &14.026 (0.042)& 14.050 (0.073)\\     
 1609$+$135  &LHS 3163      
 &14.77&14.76&14.75&14.861 (0.036) &14.779 (0.056)& 14.857 (0.109)\\     
 1625$+$093  &G138-31     
 &15.34&15.12&15.06&15.250 (0.062) &15.187 (0.103)& 15.036 (0.142)\\     
 1626$+$368  &Ross 640      
 &13.58&13.57&13.58&13.637 (0.024) &13.652 (0.034)& 13.575 (0.042)\\     
 1633$+$433  &G180-63     
 &13.95&13.76&13.73&13.991 (0.029) &13.773 (0.035)& 13.607 (0.043)\\     
 1635$+$137  &G138-47     
 &16.11&15.96&15.98:&15.929 (0.076) &15.673 (0.144)& 15.727 (0.211)\\    
 1637$+$335  &G180-65     
 &14.56&14.50&14.54&14.551 (0.031) &14.467 (0.045)& 14.424 (0.081)\\     
 1639$+$537  &GD 356        
 &14.54&14.46&14.42&14.493 (0.027) &14.479 (0.048)& 14.369 (0.085)\\     
 1655$+$215  &LHS 3254      
 &13.89&13.80&13.85&13.886 (0.026) &13.816 (0.030)& 13.863 (0.050)\\     
 1705$+$030  &G139-13     
 &14.62&14.50&14.48&14.565 (0.032) &14.499 (0.032)& 14.511 (0.078)\\     
 1716$+$020  &G19-20      
 &14.68&14.65&14.71:&14.603 (0.056) &14.534 (0.070)& 14.562 (0.109)\\    
 1733$-$544  &L270-137    
 &14.89&14.55&14.46&14.802 (0.044) &14.677 (0.084)& 14.693 (0.105)\\     
 1736$+$052  &G140-2      
 &15.62&15.56&15.49&15.682 (0.067) &15.573 (0.122)& 15.351 (0.175)\\     
 1748$+$708  &G240-72     
 &12.77&12.70&12.50&12.709 (0.021) &12.528 (0.023)& 12.507 (0.023)\\     
 1811$+$327A &G206-17     
 &15.71&15.56&15.54&15.716 (0.057) &15.675 (0.126)& 15.760 (0.201)\\     
 1811$+$327B &G206-18     
 &16.08&15.94&15.82:&16.214 (0.091) &15.953 (0.172)& 15.809 (0.213)\\    
 1818$+$126  &G141-2      
 &15.07&14.90&14.87&14.989 (0.040) &14.885 (0.069)& 14.876 (0.108)\\     
 1820$+$609  &G227-28     
 &13.96&13.73&13.65&14.075 (0.032) &13.810 (0.030)& 13.797 (0.052)\\     
 1824$+$040  &G21-15      
 &14.07&14.14&14.14&14.107 (0.032) &14.111 (0.045)& 14.225 (0.084)\\     
 1829$+$547  &G227-35     
 &14.76&14.61&14.50&14.803 (0.045) &14.478 (0.053)& 14.505 (0.078)\\     
 1831$+$197  &G184-12     
 &15.93&15.82&15.81&15.977 (0.095) &16.043 (0.184)& 15.608 (0.198)\\     
 1840$+$042  &GD 215        
 &14.53:&14.46:&14.50:&14.443 (0.050) &14.374 (0.075)& 14.651 (0.099)\\  
 1855$+$338  &G207-9      
 &14.74&14.72&14.77&14.737 (0.034) &14.769 (0.056)& 14.799 (0.124)\\     
 1917$+$386  &G125-3      
 &13.77&13.69&13.59&13.776 (0.030) &13.669 (0.032)& 13.519 (0.025)\\     
 1953$-$011  &LHS 3501      
 &13.12&13.02&13.02&13.070 (0.029) &13.029 (0.031)& 13.014 (0.040)\\     
 2002$-$110  &LHS 483       
 &15.32&15.11&15.09&15.276 (0.055) &14.995 (0.072)& 14.746 (0.105)\\     
 2011$+$065  &G24-9       
 &14.94&14.79&14.75&15.021 (0.049) &14.878 (null)& 15.090 (null)\\       
 2048$+$263  &G187-8      
 &14.12&13.83&13.79&14.100 (0.056) &13.908 (0.068)& 13.602 (null)\\      
 2054$-$050  &vB 11         
 &14.82&14.61&14.54&14.734 (0.081) &14.565 (0.134)& 14.327 (0.136)\\     
 2059$+$190  &G144-51     
 &15.52&15.36&15.34&15.642 (0.070) &15.559 (0.141)& 15.397 (0.159)\\     
 2059$+$247  &G187-16     
 &15.45&15.29&15.26&15.522 (0.057) &15.205 (0.082)& 15.061 (0.143)\\     
 2059$+$316  &G187-15     
 &14.94&14.97&14.98&14.968 (0.053) &14.927 (0.068)& 14.980 (0.115)\\     
 2105$-$820  &L24-52      
 &13.52&13.53&13.58&13.478 (0.026) &13.451 (0.033)& 13.533 (0.039)\\     
 2107$-$216  &LHS 3636      
 &15.63&15.45:&15.40:&15.688 (0.055) &15.476 (0.106)& 15.695 (0.214)\\   
 2111$+$261  &G187-32     
 &14.15&14.08&14.09&14.230 (0.036) &14.116 (0.041)& 14.095 (0.057)\\     
 2136$+$229  &G126-18     
 &15.04&14.96&15.09&15.106 (0.060) &15.055 (0.087)& 14.816 (0.116)\\     
 2140$+$207  &LHS 3703      
 &12.95&12.93&12.95&12.981 (0.021) &12.928 (0.035)& 12.922 (0.029)\\     
 2207$+$142  &G18-34      
 &14.99&14.81&14.84&14.971 (0.040) &14.782 (0.086)& 14.772 (0.098)\\     
 2246$+$223  &G67-23      
 &14.28&14.31&14.37&14.341 (0.029) &14.317 (0.047)& 14.360 (0.090)\\     
 2248$+$293  &G128-7      
 &14.24&14.01&13.94&14.316 (0.029) &13.983 (0.038)& 13.941 (0.044)\\     
 2251$-$070  &LP 701-29   
 &13.86&13.63&13.47&14.013 (0.026) &13.685 (0.036)& 13.546 (0.053)\\     
 2253$-$081  &G156-64     
 &15.59&15.47&15.36:&15.629 (0.067) &15.279 (0.086)& 15.195 (0.168)\\    
 2311$-$068  &G157-34     
 &14.98&14.93&14.90&14.951 (0.036) &14.942 (0.071)& 14.730 (0.093)\\     
 2312$-$024  &LHS 3917      
 &15.70&15.53&15.58&15.488 (0.059) &15.754 (0.170)& 14.862 (null)\\      
 2316$-$064  &LHS 542       
 &16.38&16.14&16.10&16.306 (0.092) &15.837 (0.139)& 15.200 (null)\\      
 2323$+$157  &GD 248        
 &15.06&15.08&15.06&15.051 (0.043) &14.938 (0.072)& 14.881 (0.138)\\     
 2329$+$267  &G128-72     
 &15.13&15.03&15.18&15.184 (0.041) &15.100 (0.087)& 15.030 (0.111)\\     
 2345$-$447  &ESO 292-43  
 &16.66&16.59&16.33&16.517 (0.142) &16.360 (null)& 16.332 (null)\\       
 2347$+$292  &LHS 4019      
 &14.59&14.35&14.24&14.571 (0.029) &14.345 (0.044)& 14.159 (0.065)\\     
 2352$+$401  &G171-27     
 &14.57&14.52&14.50&14.576 (0.038) &14.453 (0.061)& 14.508 (0.086)\\     
\enddata
\tablecomments{
Table 1 is available in its entirety in the electronic edition of the
{\it Astrophysical Journal}. A portion is shown here for guidance regarding
its form and content. CIT uncertainties are 5\% except
for the data marked ``:'' or ``::'', which indicate 
10\% and 20\% uncertainties, respectively. 2MASS magnitudes with
null uncertainties are lower limits.}
\end{deluxetable}

\clearpage
\clearpage
\begin{deluxetable}{lccccc}
\tabletypesize{\scriptsize}
\tablecolumns{5}
\tablewidth{0pt}
\tablecaption{Statistical Comparison of CIT and 2MASS Magnitudes}
\tablehead{
\colhead{Bandpass} &
\colhead{No. of Stars} &
\colhead{Mean} &
\colhead{Standard Deviation} &
\colhead{$\langle\sigma_{\rm 2MASS}\rangle$} &
\colhead{$\langle\sigma\rangle$\tablenotemark{a}}
}
\startdata
$J_{\rm CIT}$ - $J_{\rm 2MASS}$                   &159&$-$0.0046&0.0805&0.0502&0.0745\\
$H_{\rm CIT}$ - $H_{\rm 2MASS}$                   &157&$+$0.0180&0.1126&0.0807&0.0997\\ 
$K_{\rm CIT}$ - $K_{S\,{\rm 2MASS}}$              &143&$+$0.0247&0.1561&0.1096&0.1253\\
$J_{\rm CIT}$ - $J_{\rm 2MASS}$ (S/N $>$ 10)      &130&$-$0.0083&0.0679&0.0409&0.0662\\
$H_{\rm CIT}$ - $H_{\rm 2MASS}$ (S/N $>$ 10)      &97 &$+$0.0094&0.0675&0.0502&0.0726\\
$K_{\rm CIT}$ - $K_{S\,{\rm 2MASS}}$ (S/N $>$ 10) &49 &$+$0.0133&0.0692&0.0466&0.0697\\
\enddata
\tablenotetext{a}{Average value of $\sigma$ where for a single star, $\sigma=(\sigma_{\rm
2MASS}^2+\sigma_{\rm CIT}^2)^{1/2}$.}
\end{deluxetable}

\clearpage
\clearpage
\begin{deluxetable}{lccccccccc}
\tabletypesize{\scriptsize}
\tablecolumns{10}
\tablewidth{0pt}
\tablecaption{Sample of White Dwarfs with Predicted NIR Photometry}
\tablehead{
\colhead{WD} &
\colhead{$T_{\rm eff}$ (K)} &
\colhead{$\log g$} &
\colhead{$V$} &
\colhead{$J_{\rm pred}$} &
\colhead{$H_{\rm pred}$} &
\colhead{$K_{S\,{\rm pred}}$} &
\colhead{$J_{\rm 2MASS}$ ($\sigma_J$)} &
\colhead{$H_{\rm 2MASS}$ ($\sigma_H$)} &
\colhead{$K_{S\,{\rm 2MASS}}$ ($\sigma_K$)}}
\startdata
0023$+$388   & 10785& 8.14&15.97&15.988&15.979&16.121&13.810 (0.026) &13.268 (0.030)& 12.939 (0.033) \\
0034$-$211   & 17217:& 8.04:&14.53&14.934&15.000&15.161&11.454 (0.023) &10.884 (0.021)& 10.648 (0.026) \\
0131$-$163   & 49042& 7.81&13.96&14.667&14.808&15.000&12.966 (0.027) &12.468 (0.028)& 12.215 (0.030) \\
0145$-$257   & 25635& 7.97&14.51&15.089&15.200&15.382&12.412 (0.026) &11.830 (0.021)& 11.594 (0.023) \\
0145$-$221   & 11549& 8.14&14.85&14.965&14.974&15.122&14.923 (0.032) &14.450 (0.045)& 14.335 (0.064) \\
0205$+$133   & 58692& 7.63&15.30&16.009&16.152&16.342&12.799 (0.022) &12.198 (0.024)& 11.961 (0.020) \\
0303$-$007   & 18700& 7.97&16.00&16.439&16.517&16.679&13.164 (0.024) &12.627 (0.027)& 12.405 (0.026) \\
0347$-$137   & 12621& 8.19&14.00&14.205&14.230&14.380&12.080 (0.029) &11.540 (0.029)& 11.296 (0.023) \\
0353$+$284   & 31000:& 7.90:&11.70&12.350&12.479&12.672& 9.843 (0.023) & 9.275 (0.024)&  9.057 (0.017) \\
0429$+$176   & 13600& 8.56&13.93&14.188&14.214&14.372&10.753 (0.021) &10.161 (0.019)&  9.913 (0.017) \\
0430$+$136   & 35976& 7.90&16.45&17.126&17.263&17.456&13.533 (0.021) &12.877 (0.023)& 12.634 (0.026) \\
0628$-$020   &  6912& 8.15&15.33&14.509&14.311&14.371&10.729 (0.027) &10.144 (0.026)&  9.857 (0.024) \\
0710$+$741   & 10119& 8.11&14.97&14.881&14.851&14.986&14.692 (0.033) &14.423 (0.061)& 14.148 (0.065) \\
0812$+$478   & 60923& 7.58&15.22&15.931&16.074&16.264&14.587 (0.032) &14.165 (0.041)& 13.882 (0.047) \\
0915$+$201   & 69970& 7.33&16.64&17.354&17.498&17.687&15.721 (0.058) &15.166 (0.078)& 14.867 (0.080) \\
0950$+$139   & 94402& 9.18&16.03&16.771&16.921&17.117&16.518 (0.097) &15.945 (0.157)& 16.099 (0.258) \\
1001$+$203   & 21492& 7.97&15.35&15.849&15.944&16.110&12.640 (0.021) &12.028 (0.021)& 11.766 (0.020) \\
1013$-$050   & 60265& 7.93&14.18&14.893&15.037&15.228&10.607 (0.027) & 9.990 (0.025)&  9.770 (0.023) \\
1026$+$002   & 17183& 7.97&13.60&14.001&14.068&14.228&11.751 (0.024) &11.219 (0.027)& 10.943 (0.021) \\
1037$+$512   & 20099& 8.03&16.25&16.721&16.809&16.971&13.796 (0.024) &13.261 (0.026)& 12.972 (0.026) \\
1108$+$325   & 62950& 7.59&16.80&17.512&17.656&17.845&15.802 (0.072) &15.188 (0.079)& 15.228 (0.179) \\
1123$+$189   & 51682& 7.86&14.16&14.865&15.007&15.198&12.754 (0.023) &12.217 (0.019)& 11.990 (0.020) \\
1210$+$464   & 27667& 7.85&15.79&16.400&16.520&16.707&12.035 (0.023) &11.396 (0.021)& 11.161 (0.020) \\
1218$+$497   & 35656& 7.87&16.24&16.915&17.051&17.244&14.588 (0.038) &14.002 (0.036)& 13.837 (0.060) \\
1224$+$309   & 28824& 7.38&16.10&16.720&16.846&17.034&15.129 (0.048) &14.669 (0.068)& 14.393 (0.077) \\
1339$+$346   & 15959& 7.82&15.87&16.225&16.287&16.441&14.094 (0.027) &13.700 (0.031)& 13.591 (0.036) \\
1434$+$289   & 32795& 8.00&15.75&16.413&16.546&16.739&16.514 (0.119) &16.330 (0.203)& 15.924 (0.293) \\
1435$+$370   & 15268& 7.99&16.00&16.336&16.389&16.543&13.457 (0.024) &12.965 (0.025)& 12.746 (0.028) \\
1443$+$337   & 29763& 7.83&16.39&17.027&17.154&17.345&14.284 (0.030) &13.725 (0.030)& 13.516 (0.040) \\
1458$+$171   & 21945& 7.43&16.30&16.800&16.900&17.065&14.701 (0.031) &14.209 (0.045)& 13.847 (0.047) \\
1502$+$349   & 21339& 7.96&15.78&16.276&16.371&16.535&15.231 (0.045) &14.766 (0.061)& 14.314 (0.067) \\
1504$+$546   & 24689& 7.86&16.00&16.560&16.668&16.845&13.847 (0.025) &13.260 (0.026)& 13.001 (0.027) \\
1517$+$501   & 31100:& 7.84:&17.46&18.110&18.240&18.432&15.559 (0.060) &14.746 (0.071)& 14.157 (0.072) \\
1610$+$383   & 14450& 7.83&16.40&16.701&16.749&16.898&14.437 (0.034) &13.807 (0.036)& 13.521 (0.042) \\
1619$+$525   & 18041& 7.90&15.81&16.232&16.306&16.467&14.168 (0.032) &13.545 (0.035)& 13.425 (0.042) \\
1619$+$414   & 14091& 7.93&16.80&17.087&17.129&17.279&13.937 (0.021) &13.311 (0.029)& 13.025 (0.027) \\
1622$+$323   & 68277& 7.56&16.33&17.045&17.189&17.379&14.633 (0.029) &13.963 (0.031)& 13.773 (0.039) \\
1631$+$781   & 44931& 7.76&13.38&14.076&14.216&14.408&10.975 (0.021) &10.398 (0.021)& 10.164 (0.014) \\
1639$+$153   &  7482& 8.42&15.70&15.032&14.869&14.948&15.073 (0.042) &14.979 (0.087)& 15.060 (0.128) \\
1643$+$143   & 26849& 7.91&15.64&16.239&16.355&16.540&12.732 (0.024) &12.125 (0.031)& 11.957 (0.024) \\
1711$+$668   & 53751& 8.47&17.00&17.711&17.854&18.048&15.120 (0.043) &14.457 (0.057)& 14.211 (0.087) \\
1717$-$345   & 12700:& 7.75:&16.38&16.588&16.620&16.762&12.870 (0.039) &12.208 (0.060)& 11.940 (0.054) \\
2151$-$015   &  9137& 8.21&14.41&14.122&14.049&14.168&12.452 (0.029) &11.778 (0.022)& 11.414 (0.027) \\
2256$+$249   & 22151& 7.82&13.64&14.150&14.249&14.416&11.675 (0.020) &11.180 (0.025)& 10.915 (0.018) \\
2257$+$162   & 27556& 8.33&16.14&16.756&16.873&17.061&15.439 (0.054) &15.088 (0.074)& 14.736 (0.108) \\
2317$+$268   & 31460& 7.70&16.30&16.951&17.082&17.275&14.609 (0.033) &14.074 (0.036)& 13.783 (0.050) \\
2336$-$187   &  7882& 7.82&15.60&15.035&14.896&14.987&15.057 (0.040) &14.939 (0.063)& 14.681 (0.093) \\
\enddata
\tablecomments{
Uncertainties of the atmospheric parameters are 1.2\% in $\Te$ and
0.038 dex in $\logg$. The $V$ magnitudes are from various sources in
the literature. The objects marked with a colon are contaminated
by the companion in the visible and the uncertainties are correspondingly larger.}
\end{deluxetable}

\clearpage
\clearpage
\begin{deluxetable}{lccccccccc}
\tabletypesize{\scriptsize}
\tablecolumns{10}
\tablewidth{0pt}
\tablecaption{Sample of Massive White Dwarfs}
\tablehead{
\colhead{WD} &
\colhead{$T_{\rm eff}$ (K)} &
\colhead{M/$M_{\sun}$} &
\colhead{$V$} &
\colhead{$J_{\rm pred}$} &
\colhead{$H_{\rm pred}$} &
\colhead{$K_{\rm pred}$} &
\colhead{$J_{\rm 2MASS}$ ($\sigma$$_{\rm J}$)} &
\colhead{$H_{\rm 2MASS}$ ($\sigma$$_{\rm H}$)} &
\colhead{$K_S$$_{\rm 2MASS}$ ($\sigma$$_{\rm K}$)}}
\startdata
0033+016& 10984& 1.11&15.61&15.638&15.632&15.679&15.650 (0.057) &15.522 (0.090)& 16.119 (0.303) \\
0052+226&  9652& 1.05&16.16&15.966&15.918&15.947&16.021 (0.077) &16.109 (0.162)& 15.522 (0.212) \\
0052$-$147& 25683& 0.80&15.12&15.715&15.832&15.912&15.724 (0.061) &15.532 (0.109)& 15.457 (null) \\
0101+059& 14191& 0.83& --&--&--&--&16.214 (0.089) &16.387 (0.210)& 17.182 (null) \\
0143+216&  9292& 0.92&15.05&14.792&14.732&14.755&14.784 (0.036) &14.812 (0.060)& 14.676 (0.077) \\
0213+396&  9323& 0.96&14.54&14.287&14.227&14.250&14.304 (0.030) &14.223 (0.052)& 14.144 (0.049) \\
0231$-$054& 13552& 1.02&14.28&14.545&14.583&14.643&14.540 (0.033) &14.558 (0.052)& 14.659 (0.101) \\
0232+525& 16738& 0.80&13.75&14.150&14.220&14.283&14.218 (0.031) &14.261 (0.053)& 14.497 (0.072) \\
0346$-$011& 41185& 1.27&14.01&14.728&14.867&14.957&14.747 (0.030) &14.863 (0.038)& 15.120 (0.138) \\
0429+176& 13600& 0.97&13.93&14.192&14.231&14.290&10.753 (0.021) &10.161 (0.019)&  9.913 (0.017) \\
0532$-$560& 11556& 0.92&16.00&16.110&16.125&16.176&16.023 (0.085) &15.882 (0.185)& 16.348 (null) \\
0558+165& 16199& 0.81&15.69&16.071&16.137&16.199&16.004 (0.069) &16.189 (0.161)& 15.605 (null) \\
0644+025&  7242& 1.00&15.71&14.990&14.820&14.793&14.868 (0.045) &14.757 (0.069)& 14.576 (0.103) \\
0701$-$587& 13696& 0.91&14.46&14.731&14.773&14.831&14.844 (0.035) &14.856 (0.074)& 15.067 (0.150) \\
0730+487& 14311& 0.91&14.96&15.264&15.311&15.370&15.143 (0.045) &15.191 (0.094)& 15.395 (0.171) \\
0743+442& 14501& 0.84&14.87&15.183&15.234&15.292&15.230 (0.045) &15.403 (0.104)& 15.179 (0.131) \\
0827+328&  7508& 0.96&15.73&15.077&14.921&14.903&14.985 (0.044) &14.964 (0.076)& 14.865 (0.121) \\
0930+294&  8362& 0.98&15.98&15.525&15.416&15.420&15.588 (0.066) &15.399 (0.106)& 15.284 (0.150) \\
0947+325& 22055& 0.82&15.43&15.957&16.060&16.126&16.011 (0.073) &16.127 (0.171)& 16.503 (null) \\
0950+139& 94402& 1.36&16.03&16.789&16.939&17.029&16.518 (0.097) &15.945 (0.157)& 16.099 (0.258) \\
1038+633& 24447& 0.86&15.15&15.725&15.836&15.912&15.723 (0.080) &15.748 (0.188)& 15.389 (null) \\
1049$-$158& 18989& 0.83&14.36&14.820&14.907&14.970&14.789 (0.047) &14.818 (0.060)& 15.116 (0.164) \\
1052+273& 23103& 0.86&14.12&14.664&14.771&14.841&14.619 (0.029) &14.674 (0.048)& 14.784 (0.076) \\
1058$-$129& 24311& 1.06&15.75&16.328&16.437&16.513&15.520 (0.054) &15.689 (0.118)& 15.437 (0.219) \\
1102+748& 19712& 0.84&14.97&15.448&15.539&15.601&15.556 (0.059) &15.487 (0.116)& 15.552 (0.228) \\
1120+439& 26950& 0.85&15.81&16.428&16.549&16.632&16.053 (0.077) &15.977 (0.158)& 16.200 (0.362) \\
1134+300& 21276& 0.96&12.52&13.036&13.134&13.199&12.993 (0.024) &13.105 (0.031)& 13.183 (0.028) \\
1159$-$098&  9536& 1.10&15.90&15.682&15.628&15.654&15.555 (0.056) &15.480 (0.091)& 15.384 (0.186) \\
1236$-$495& 11748& 1.10&13.80&13.923&13.935&13.989&13.806 (0.024) &13.815 (0.036)& 13.907 (0.062) \\
1237$-$028& 10236& 0.97&15.97&15.888&15.865&15.903&15.971 (0.068) &15.922 (0.138)& 15.754 (0.250) \\
1257+278&  8733& 0.81&15.41&15.040&14.954&14.966&15.132 (0.046) &14.977 (0.076)& 14.986 (0.089) \\
1304+227& 10444& 0.87& --  & --& --& --&16.413 (0.112) &16.601 (0.288)& 16.498 (null) \\
1310+583& 10555& 0.80&14.09&14.070&14.063&14.106&14.016 (0.028) &14.004 (0.045)& 14.081 (0.073) \\
1334$-$160& 18667& 0.80&15.34&15.792&15.876&15.939&15.532 (0.053) &15.553 (0.103)& 15.733 (0.295) \\
1446+286& 22891& 0.89&14.54&15.086&15.192&15.261&15.172 (0.044) &15.269 (0.113)& 15.537 (0.251) \\
1452+553& 27636& 0.82& --&--&--&--&16.642 (0.118) &16.606 (0.235)& 16.720 (null) \\
1459+347& 21516& 0.92&15.74&16.259&16.359&16.424&16.402 (0.101) &16.327 (0.207)& 15.645 (null) \\
1515+668& 10317& 0.86&15.33&15.268&15.251&15.291&15.295 (0.053) &15.240 (0.112)& 15.180 (0.198) \\
1525+257& 22291& 0.80&15.65&16.181&16.286&16.352&16.258 (0.098) &16.101 (0.177)& 16.022 (null) \\
1531$-$022& 18617& 0.88&14.00&14.453&14.535&14.600&14.395 (0.040) &14.484 (0.053)& 14.618 (0.101) \\
1554+322& 30497& 0.85& --&--&--&--&16.561 (0.115) &17.617 (null)& 16.016 (0.241) \\
1609+135&  9321& 1.01&15.11&14.854&14.793&14.816&14.861 (0.036) &14.779 (0.056)& 14.857 (0.109) \\
1625+093&  6870& 0.88&16.14&15.318&15.125&15.086&15.250 (0.062) &15.187 (0.103)& 15.036 (0.142) \\
1636+057&  8537& 1.07&16.46&16.040&15.941&15.948&16.057 (0.081) &15.897 (0.147)& 15.919 (null) \\
1639+153&  7482& 0.86&15.68&15.021&14.863&14.844&15.073 (0.042) &14.979 (0.087)& 15.060 (0.128) \\
1647+591& 12258& 0.81&12.23&12.412&12.444&12.497&12.425 (0.021) &12.463 (0.021)& 12.522 (0.030) \\
1711+668& 53751& 0.96&17.00&17.728&17.872&17.960&15.120 (0.043) &14.457 (0.057)& 14.211 (0.087) \\
1840+042&  8925& 0.81&14.79&14.461&14.384&14.400&14.443 (0.050) &14.374 (0.075)& 14.651 (0.099) \\
1855+338& 11958& 0.83&14.64&14.795&14.820&14.874&14.737 (0.034) &14.769 (0.056)& 14.799 (0.124) \\
2039$-$682& 15855& 0.89&13.41&13.781&13.842&13.904&13.729 (0.026) &13.806 (0.039)& 13.800 (0.050) \\
2051$-$208& 20512& 1.23&15.06&15.570&15.659&15.725&15.590 (0.050) &15.669 (0.112)& 15.839 (0.226) \\
2059+190&  6980& 0.86&16.38&15.589&15.402&15.366&15.642 (0.070) &15.559 (0.141)& 15.397 (0.159) \\
2124+550& 13341& 0.82&14.70&14.952&14.993&15.049&14.987 (0.053) &14.957 (0.090)& 14.905 (0.162) \\
2205$-$139& 25263& 0.81&15.08&15.668&15.783&15.861&15.648 (0.067) &15.610 (0.114)& 15.582 (0.217) \\
2220+133& 22675& 0.88&15.60&16.141&16.246&16.315&16.264 (0.112) &16.046 (0.180)& 15.367 (null) \\
2246+223& 10647& 1.10&14.39&14.368&14.354&14.397&14.341 (0.029) &14.317 (0.047)& 14.360 (0.090) \\
2313+682&  8977& 0.83&16.18&15.861&15.787&15.804&15.908 (0.098) &15.766 (0.175)& 15.387 (0.210) \\
\enddata
\tablecomments{
2MASS magnitudes with null uncertainties are lower limits. Mean uncertainties of effective temperatures and masses are 1.2\% and 0.03 $M_{\sun}$, respectively.
}
\end{deluxetable}

\clearpage

\figcaption[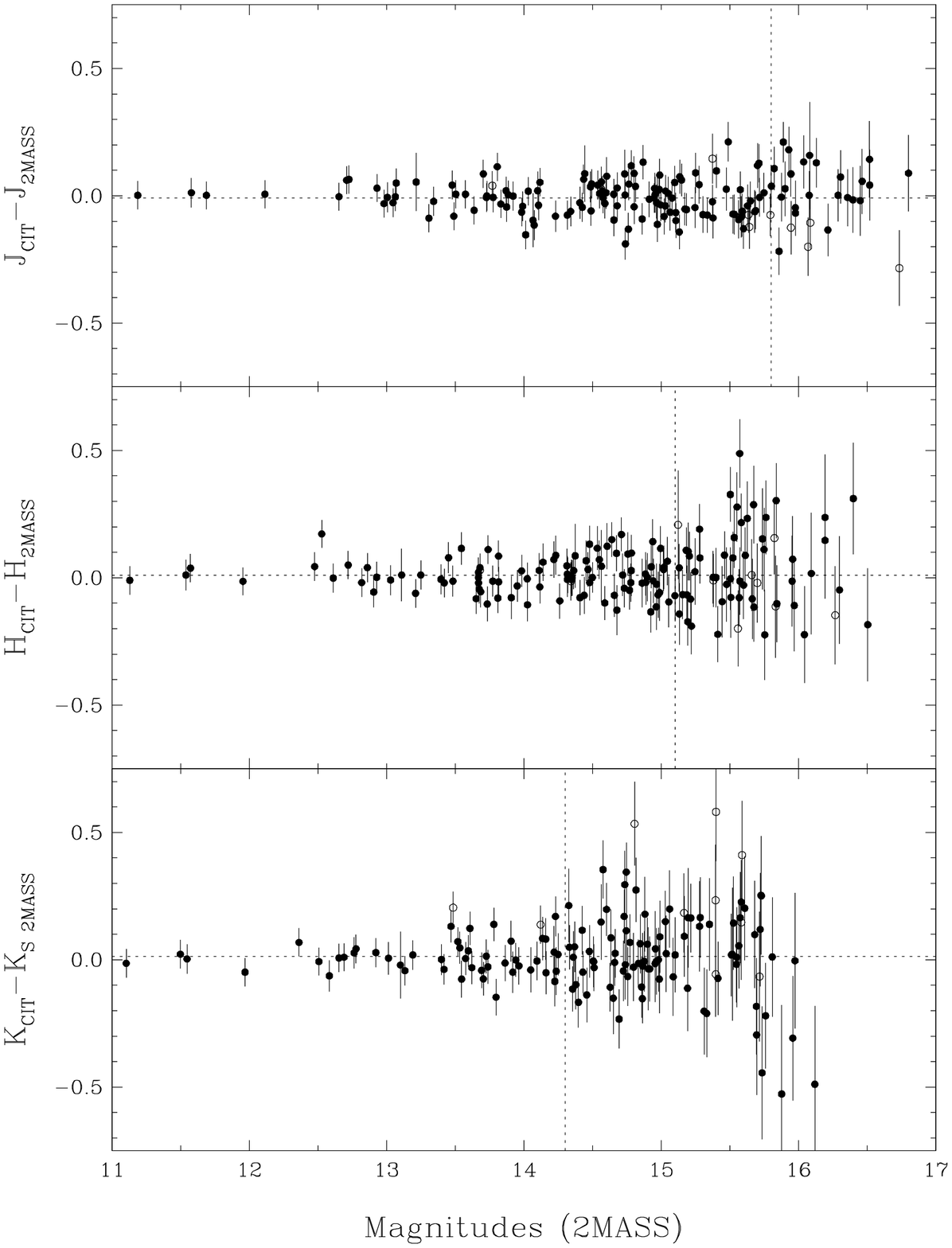] 
{Differences in magnitudes between the infrared CIT and 2MASS
photometric systems for each individual filter as a function of the
2MASS magnitude for our common sample of 160 cool white dwarfs. The
error bars represent the combined quadratic uncertainties of both
photometric data sets. The horizontal dotted lines indicate the
mean magnitude differences between both data
sets.  Objects located on the left side of the vertical dotted lines
meet the PSC level~1 requirements (S/N$\,>10$), which correspond to
$J < 15.8$, $H < 15.1$, and $K_S < 14.3$. The ten objects represented
by open circles are discussed in the text and in Figure
\ref{fg:f3}.\label{fg:f1}}

\figcaption[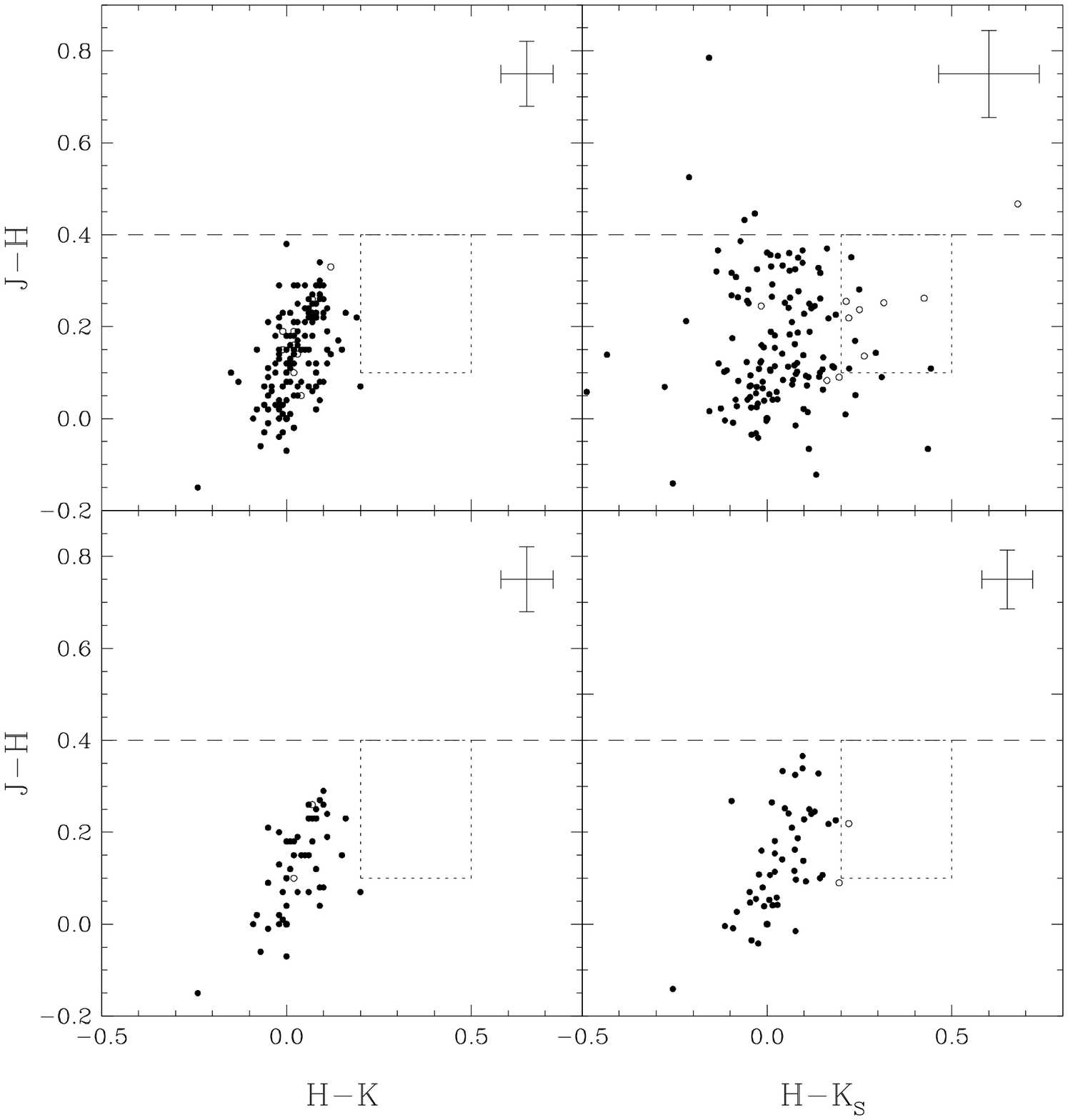] 
{{\it Top:} $(J-H)$ vs.~$(H-K/K_S)$ two-color diagrams for 143 cool
white dwarfs taken from Table 1 and detected by 2MASS in all
three bands. The left and right panels correspond to the CIT 
and 2MASS magnitudes, respectively.  The error bars
indicate the mean uncertainties of each data set.
{\it Bottom:} Same as the top panels but for the 49 white
dwarfs satisfying the level~1 requirements. The region above the dashed
line and that defined by the dotted rectangle correspond to the color
criteria defined by \citet{wachter03} for selecting binary candidates
and tentative binary candidates, respectively. The ten objects shown
by open circles are discussed in the text and in Figure
\ref{fg:f3}. \label{fg:f2}}

\figcaption[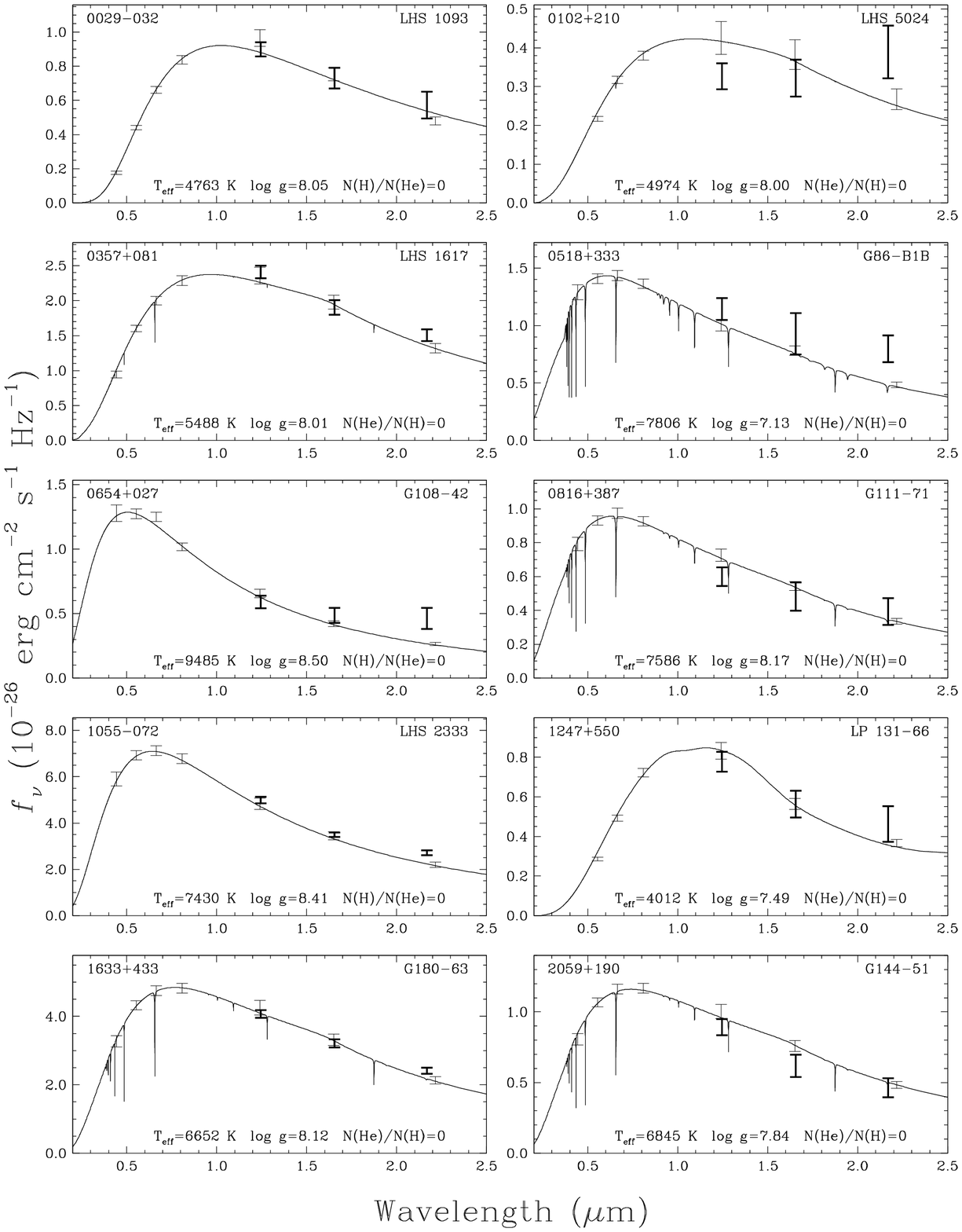]
{Fits to the optical $BVRI$ and infrared $JHK$ CIT photometric energy
distributions ({\it thin error bars}) for ten objects from our cool
white dwarf sample. The atmospheric parameters obtained from a fit to
the observed energy distribution are given in each panel and the
corresponding monochromatic model fluxes are shown by the solid line.
For clarity, we do not plot the model fluxes averaged over the filter
bandpasses and used in the fitting procedure as they coincide almost
perfectly with the monochromatic fluxes. Also shown by thick
error bars are the corresponding 2MASS fluxes. \label{fg:f3}}

\figcaption[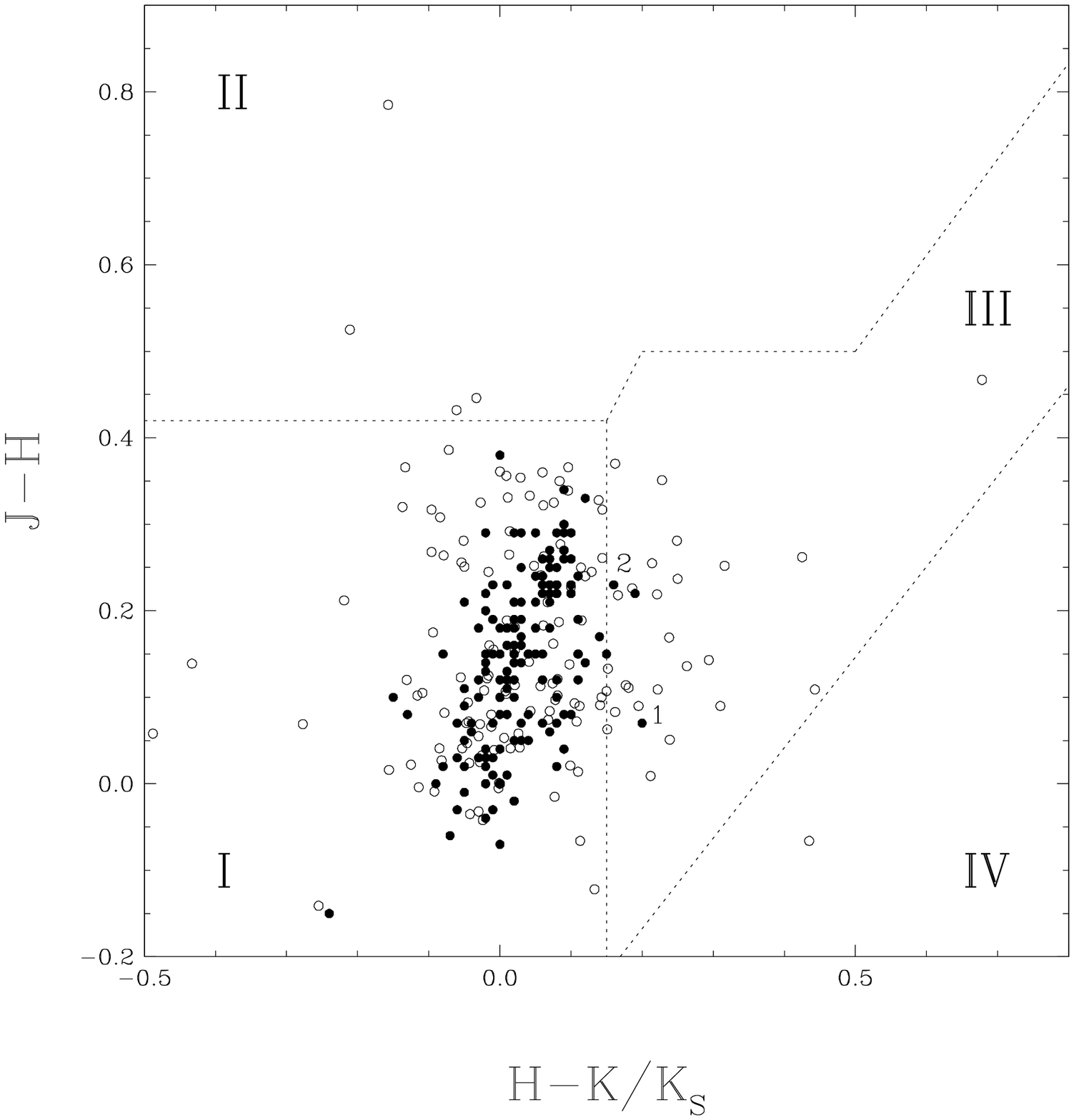]
{Same as Figure \ref{fg:f2} but with the four
regions defined by \citet[][see text]{wellhouse05}. The filled and
open circles correspond to the CIT and 2MASS colors, respectively. The
two objects with CIT data labeled in the figure and discussed
in the text are (1) 1748$+$708 and (2) 2251$-$070. \label{fg:f4}}

\figcaption[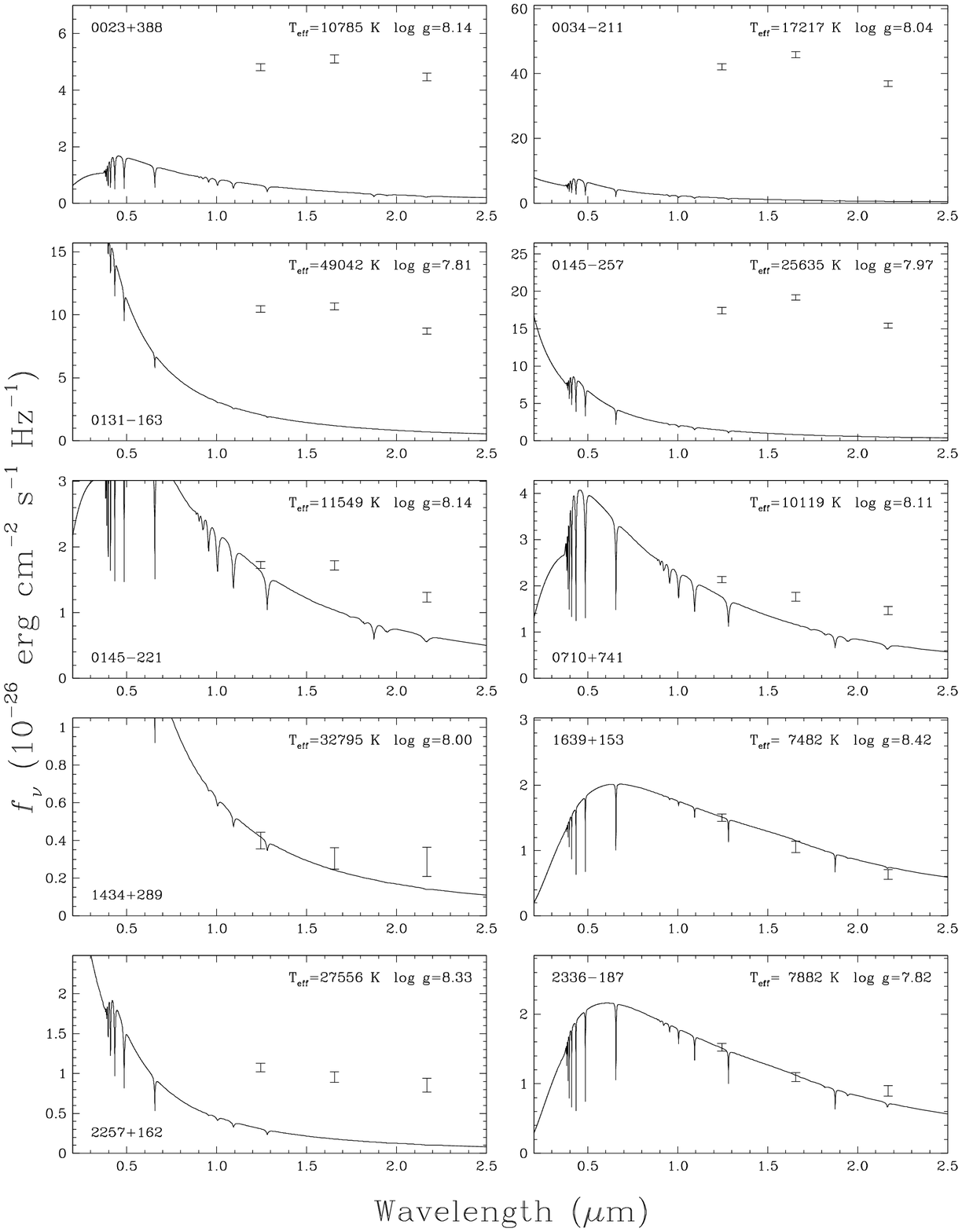]
{Observed 2MASS fluxes ({\it error bars}) for several binary and
tentative binary candidates from \citet{wachter03} compared with the
predictions of model atmospheres ({\it solid lines}) normalized at $V$. The atmospheric
parameters derived from the spectroscopic method are given in each
panel.\label{fg:f5}}

\figcaption[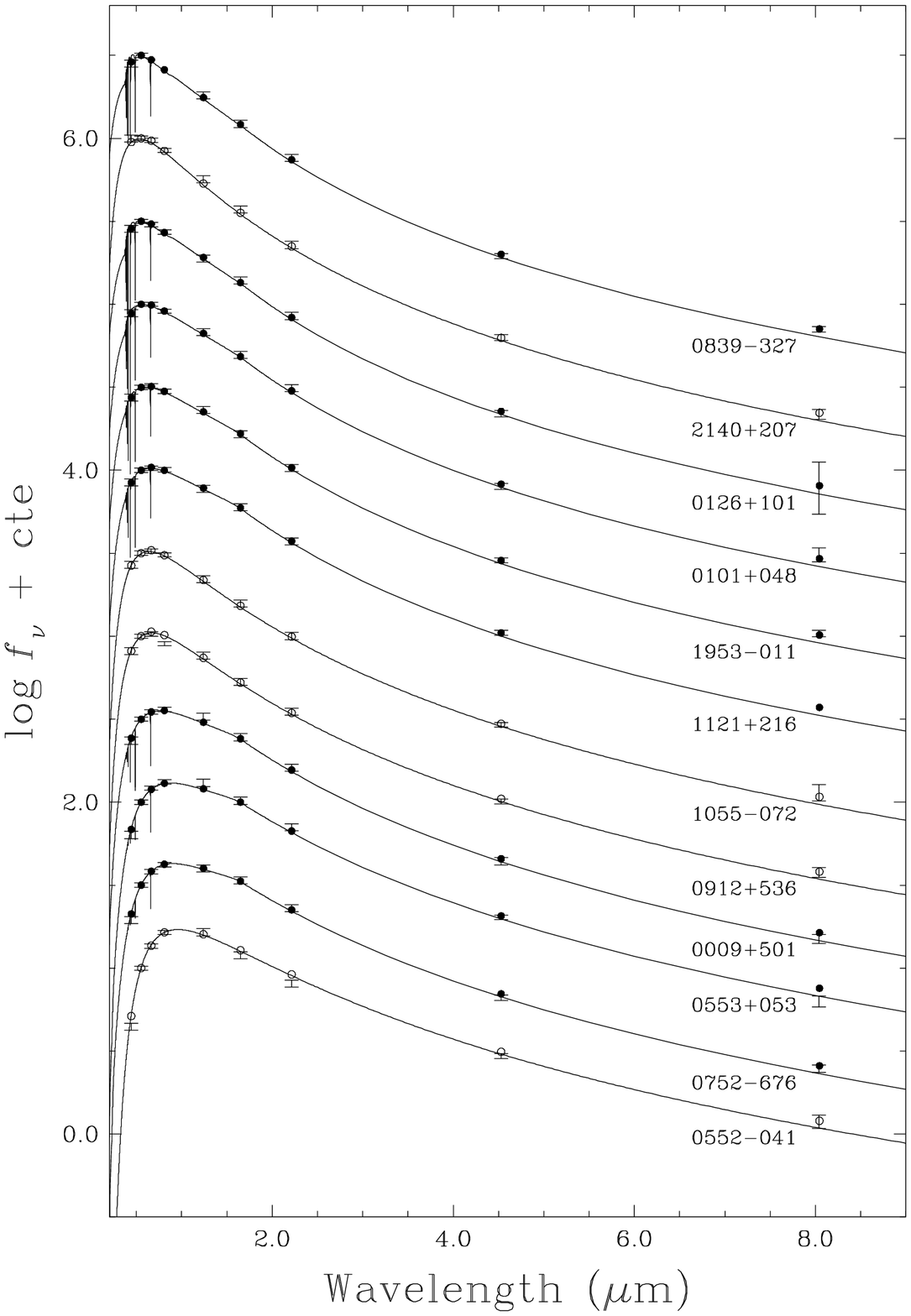] 
{Fits to the energy distribution of white dwarfs from the sample of
\citet{kilic06b}. The observed $BVRI$ and $JHK$ (CIT) fluxes along with the
4.5 and $8~\mu$m {\it Spitzer} fluxes are shown by error bars.  The
flux scale is logarithmic and each star is shifted vertically by a
constant for clarity. The model monochromatic fluxes are shown by
solid lines while the fluxes averaged over the filter bandpasses are
indicated by filled and open circles for pure hydrogen and pure helium
atmospheric compositions, respectively. \label{fg:f6}}

\figcaption[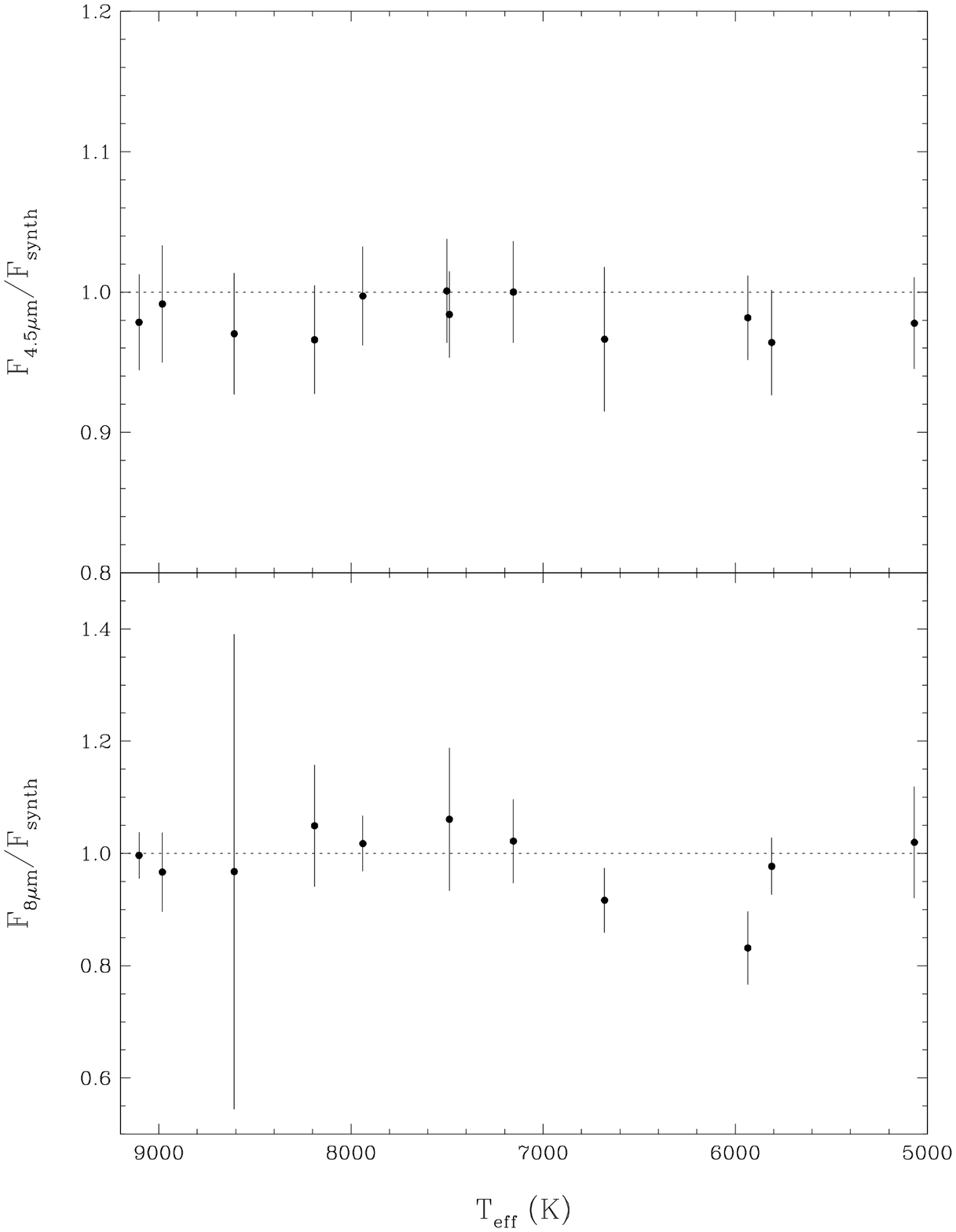] 
{The ratio of observed to predicted {\it Spitzer} fluxes for 12 objects from
the sample of \citet{kilic06b} as a function of effective
temperature. The predicted fluxes and $\Te$ values are obtained from
simultaneous fits to the $BVRIJHK$ and 4.5 and $8~\mu$m photometric
data. For Ross 627 (1121$+$216), only the {\it Spitzer} $4.5~\mu$m flux is
used since the $8~\mu$m flux is affected by a nearby star.\label{fg:f7}}

\figcaption[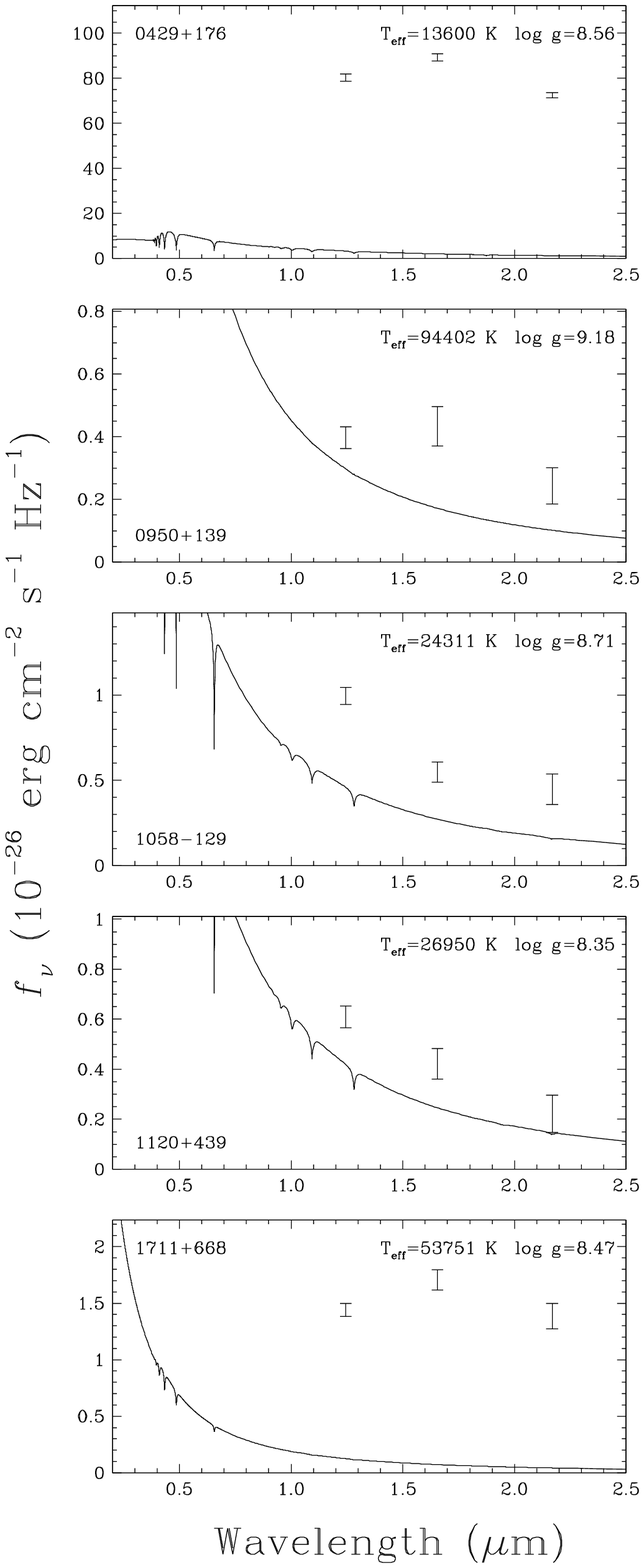] 
{Same as Fig.~\ref{fg:f5} but for the five white dwarfs discussed in
\S~5.\label{fg:f8}}

\figcaption[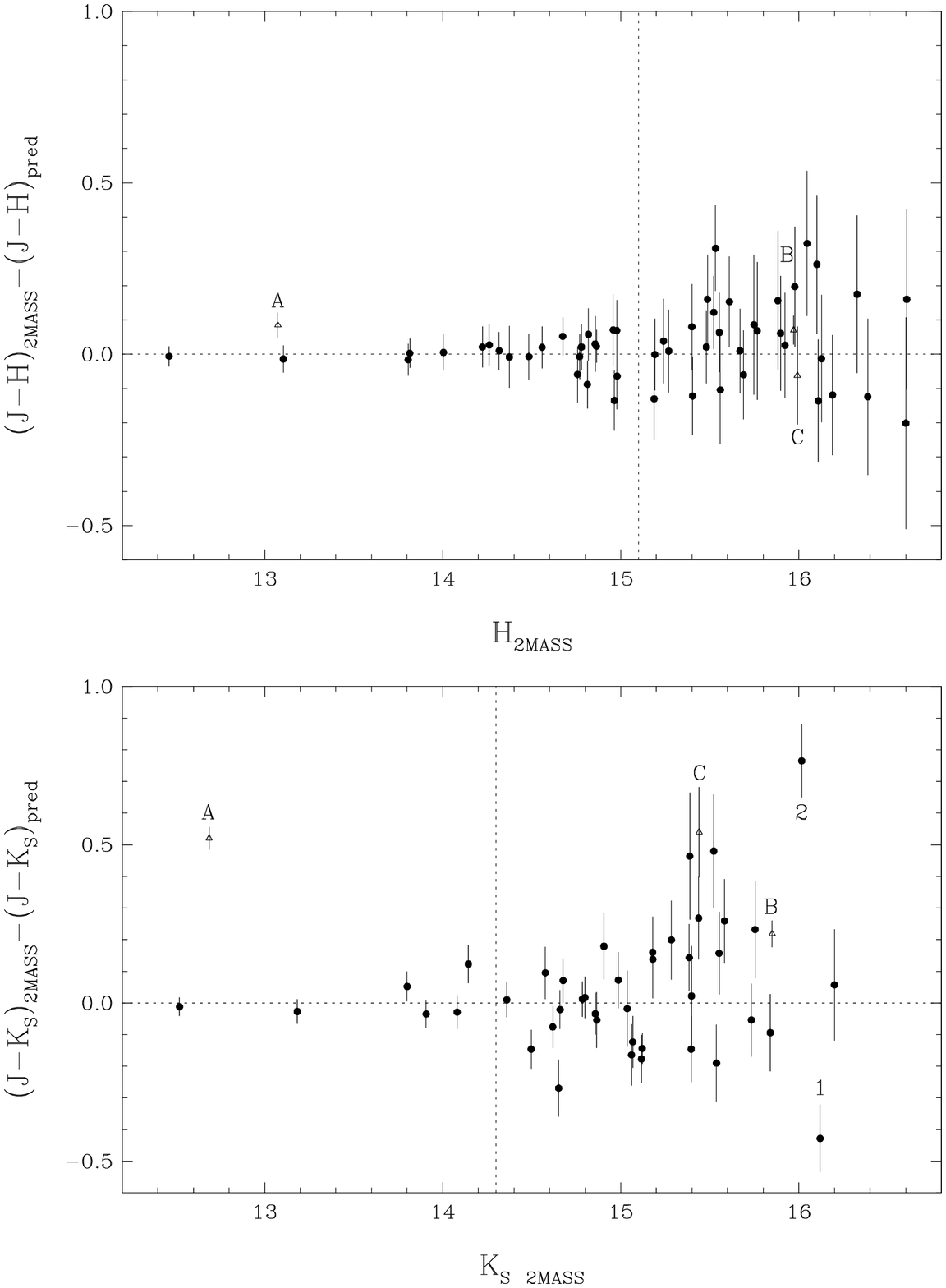] 
{Differences between the 2MASS observed and predicted $(J-H)$ and
$(J-K_S)$ color indices for our sample of massive white dwarfs (Table 4) 
as a function of $H$ and $K_S$, respectively. In the upper
panel, 53 objects are detected at both $J$ and $H$, while
in the lower panel 42 objects are detected at both $J$ and $K_s$.
The uncertainties are from the 2MASS PSC. Known white
dwarfs (not necessarily massive) with a circumstellar disk are shown
by triangles and correspond to (A) G29-38, (B) GD~362 and (C) GD~56;
the observed colors for G29-38 and GD 56 are from 2MASS and for GD 362 
from \citet{becklin05}. Also identified in the figure and discussed
in the text are (1) G1-7 and (2) CBS 413.
Left of the dotted vertical lines are objects for which the 2MASS
level-1 requirements are satisfied for the corresponding color
index. The horizontal dotted lines represent identical values of
observed and theoretical color indices. \label{fg:f9}}

\clearpage

\begin{figure}[p]
\plotone{f1.eps}
\begin{flushright}
Figure \ref{fg:f1}
\end{flushright}
\end{figure}

\clearpage

\begin{figure}[p]
\plotone{f2.eps}
\begin{flushright}
Figure \ref{fg:f2}
\end{flushright}
\end{figure}

\clearpage

\begin{figure}[p]
\plotone{f3.eps}
\begin{flushright}
Figure \ref{fg:f3}
\end{flushright}
\end{figure}

\clearpage

\begin{figure}[p]
\plotone{f4.eps}
\begin{flushright}
Figure \ref{fg:f4}
\end{flushright}
\end{figure}

\clearpage

\begin{figure}[p]
\plotone{f5.eps}
\begin{flushright}
Figure \ref{fg:f5}
\end{flushright}
\end{figure}

\clearpage

\begin{figure}[p]
\plotone{f6.eps}
\begin{flushright}
Figure \ref{fg:f6}
\end{flushright}
\end{figure}

\clearpage

\begin{figure}[p]
\plotone{f7.eps}
\begin{flushright}
Figure \ref{fg:f7}
\end{flushright}
\end{figure}

\clearpage

\begin{figure}[p]
\plotone{f8.eps}
\begin{flushright}
Figure \ref{fg:f8}
\end{flushright}
\end{figure}

\clearpage

\begin{figure}[p]
\plotone{f9.eps}
\begin{flushright}
Figure \ref{fg:f9}
\end{flushright}
\end{figure}

\end{document}